\documentclass[aps,prl,showpacs,twocolumn,nofootinbib,preprintnumbers]{revtex4-2}
\usepackage{bbm}
\usepackage{mathrsfs}
\usepackage{epsfig}
\usepackage{soul,xcolor}
\usepackage{graphicx}
\usepackage{amsfonts}
\usepackage{amsthm}
\usepackage[figuresright]{rotating}
\usepackage{amssymb}
\usepackage{amsmath}
\usepackage{dcolumn}
\usepackage{physics}
\usepackage{float}
\usepackage{bm}
\usepackage{physics}
\usepackage{verbatim}
\usepackage{braket}
\usepackage[normalem]{ulem}
\usepackage[ruled,vlined,linesnumbered]{algorithm2e}
\usepackage{setspace}
\usepackage{lipsum}
\usepackage[colorlinks,linkcolor=blue,anchorcolor=blue,citecolor=blue,urlcolor=blue]{hyperref}
\newtheorem{theorem}{Theorem}

\newcommand\blfootnote[1]{%
  \begingroup
  \renewcommand\thefootnote{}\footnote{#1}%
  \addtocounter{footnote}{-1}%
  \endgroup
}

\begin{document}

\title{Hamiltonian Learning at Heisenberg Limit for Hybrid Quantum Systems}
\author{Lixing Zhang$^{1}$, Ze-Xun Lin$^{2}$, Prineha Narang$^{2,3}$, Di Luo$^{3*}$\thanks{email: \url{diluo@ucla.edu}}}

\affiliation{$^{1}$\mbox{Department of Chemistry and Biochemistry, University of California Los Angeles, Los Angeles, CA 90095, USA}\\
$^{2}$\mbox{Division of Physical Sciences, College of Letters and Science, University
of California, Los Angeles, 90095, California, USA}\\
$^{3}$\mbox{Department of Electrical and Computer Engineering, University of California Los Angeles, Los Angeles, CA 90095, USA} \\
}
\begin{abstract}
Hybrid quantum systems with different particle species are fundamental in quantum materials and quantum information science. 
In this work, we establish a rigorous theoretical framework proving that, given access to an unknown spin-boson type Hamiltonian, our algorithm achieves Heisenberg-limited estimation for all coupling parameters up to error $\epsilon$ with a total evolution time $\mathcal{O}(\epsilon^{-1})$ using only $\mathcal{O}({\rm polylog}(\epsilon^{-1}))$ measurements. It is also robust against small state preparation and measurement errors. In addition, we provide an alternative algorithm based on distributed quantum sensing, which significantly reduces the evolution time per measurement. To validate our method, we demonstrate its efficiency in hybrid Hamiltonian learning and spectrum learning, with broad applications in AMO, condensed matter and high energy physics. Our results provide a scalable and robust framework for precision Hamiltonian characterization in hybrid quantum platforms.

\end{abstract}

\maketitle

\textit{Introduction---.} The study of hybrid quantum systems of different particle species is essential for both understanding fundamental interactions in nature and advancing quantum engineering.
In condensed matter physics, electron-phonon interaction historically played a pivotal role in the breakthrough understanding of BCS superconductivity~\cite{bardeen1957theory}. More broadly, fermion interactions with gauge bosons have been a unifying theme in quantum field theory~\cite{feynman2018space,schwinger1951theory,yang1954conservation}. In particular, (2+1)-dimensional quantum electrodynamics (QED$_3$) provides valuable insights into the cavity-enhanced fractional quantum Hall effect~\cite{FQHE}, topological quantum matter~\cite{TI}, and quantum spin liquids~\cite{savary2016quantum}.
In quantum information science, hybrid quantum systems facilitate information transfer between superconducting qubits and infrared photons in quantum transduction~\cite{QT} and enable long-distance communication via optical fiber-based quantum networks~\cite{Metro_scale}. Recently, hybrid qauntum computing platforms based on Rydberg atoms, ion traps, and superconducting qubit-oscillator~\cite{GKP, HybridQC1, HybridQC2} have emerged, providing novel avenues for fault-tolerant quantum simulation and algorithmic advancements.\blfootnote{*Email: ~\url{diluo@ucla.edu}}

To bridge theory and experiment, the characterization of Hamiltonian is crucial, which helps to derive accurate theoretical models~\cite{HL1}, calibrate quantum devices~\cite{HL2}, and design error-mitigation algorithms~\cite{HL3}. This necessity has led to the field of Hamiltonian learning, which reconstructs a quantum system’s Hamiltonian from experimental measurements. According to the central limit theorem,  achieving a root-mean-square error (RMSE) of $\epsilon$ in any measured quantities requires a scaling of $\mathcal{O}(\epsilon^{-2})$  in total evolution time and number of measurements, defining the standard quantum limit (SQL). On the other hand, as predicted by the uncertainty principle, the optimal scaling in total evolution time is $\mathcal{O}(\epsilon^{-1})$, known as the Heisenberg limit. Recently, Heisenberg limit hamiltonian learning is achieved for qubit systems~\cite{Hsin2023,Ainesh2024,Hu2025} and further generalized to bosonic and fermionic hamiltonians~\cite{Haoya2023,Arjun2024}. Despite the above advancement, efficient learning for hybrid Hamiltonians remains an open challenge. The coupling between discrete and continuous degrees of freedom (DOFs) in hybrid systems leads to complex dynamic phenomena across different parameter regimes. However, the fundamental asymmetry between discrete states and continuous modes presents significant theoretical and computational hurdles, as traditional methods struggle to capture both quantization and infinite-dimensional dynamics. Even for the simplest hybrid system, a general solution was elusive until~\cite{Rabi}, highlighting the greater complexity of hybrid Hamiltonian learning.

\begin{figure*}[t]
\centerline{\includegraphics[width=180mm]{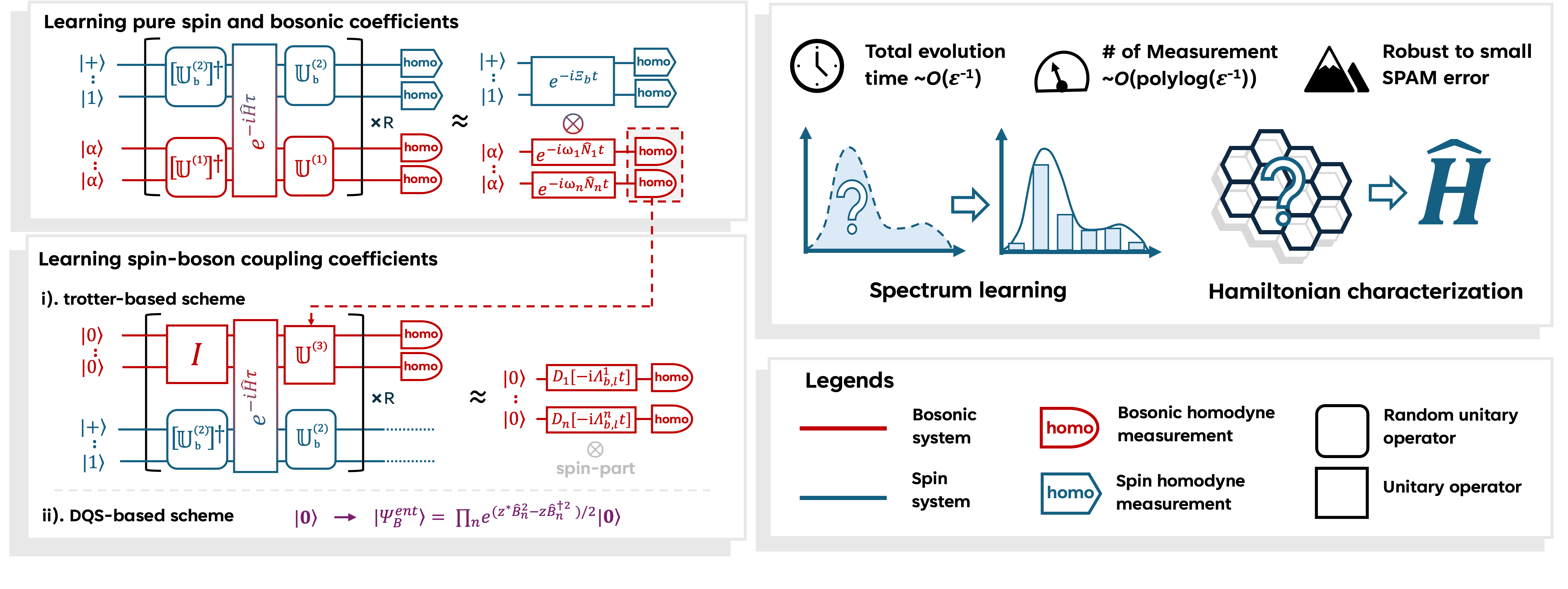}}
\caption{Hybrid Quantum systems Hamiltonian learning. Left panel: Schematic of the learning protocol presented in this work. Right panel: Properties and applications of the main algorithm. }

\label{flowchart}
\end{figure*}

In this paper, we present the first algorithm that learns hybrid Hamiltonians of spin-boson at the Heisenberg limit. In learning all Hamiltonian coefficients to an RMSE of $\epsilon$, the proposed algorithm only uses $\mathcal{O}(\epsilon^{-1})$ total evolution time and $\mathcal{O}({\rm polylog}(\epsilon^{-1}))$ measurements, while maintaining robustness against small state preparation and measurement (SPAM) errors. Besides the main algorithm, we also provide an alternative scheme based on distributed quantum sensing (DQS), which significantly reduces the maximum evolution time per measurement. We numerically verify the scaling of our approaches and provide concrete examples that demonstrate its applicability in near-term quantum devices and spectrum learning problems.

\textit{Hybrid Hamiltonian learning}---We consider Hamiltonian that has the following form:
\begin{eqnarray}\label{gen_H}
{\hat H} = {\hat H}_{\rm spin} + {\hat H}_{\rm boson} + {\hat H}_{\rm int}
\end{eqnarray}
where ${\hat H}_{\rm spin} = \sum_a \xi_a \hat{E}_a$, ${\hat H}_{\rm boson} = \sum_n \omega_n {\hat b}_n^\dagger {\hat b}_n$ and ${\hat H}_{\rm int} = \sum_{n,a} \lambda_a^n\hat{E}_a{\hat X}_n$. For the $n^{th}$ bosonic mode, ${\hat b}_n^\dagger ({\hat b}_n)$ is the creation (annihilation) operator, whereas ${\hat X}_n \equiv ({\hat b}_n^\dagger + {\hat b}_n)/\sqrt{2}$ is the position operator. $\hat{E}_{a}$ is a multi-qubit Pauli string, and the spin part of ${\hat H}_{\rm int}$ is assumed to be $k$-local, i. e. $|{\rm supp}(E_{a})| \le k$. For example, a 5-qubit spin Hamiltonian with $k = 3$, $\hat{E}_{a}$ could be $XIZYI$. The form of ${\hat H}$ covers a wide range of models, including the Holstein model and the Su–Schrieffer–Heeger (SSH) model. Explicit expressions for these models are provided in ~\cite{supmat}.

The algorithm proposed in this work achieves the following properties:
\begin{theorem}\label{th1}
Given a unitary dynamics access to arbitrary Hamiltonian in the form of Eq.~(\ref{gen_H}), there is an algorithm $\mathcal{A}$ that can estimate $\xi_a$, $\lambda_a^n$, $\eta_a^n$ and $\omega_n$ up to a RMSE $\epsilon$ such that:
\begin{enumerate}
  \item $\mathcal{A}$ takes a total evolution time of $T \sim \mathcal{O}(\epsilon^{-1})$ to measure all of the coefficients.
  \item $\mathcal{A}$ uses $\mathcal{O}({\rm polylog(\epsilon^{-1}) )}$ measurements to learn all of the coefficients.
  \item $\mathcal{A}$ is robust under small SPAM error.
\end{enumerate}
\end{theorem}
To achieve Theorem~\ref{th1}, we first cancel ${\hat H}_{\rm int}$ via random unitary transformation (RUT). This allow us to learn ${\hat H}_{\rm spin}$ and ${\hat H}_{\rm boson}$ independently. Compared to~\cite{Hsin2023}, our algorithm for ${\hat H}_{\rm spin}$ works without the low-intersection assumption, making it scalable for general $k$-local spin Hamiltonians. To learn ${\hat H}_{\rm int}$, we use the learnt coefficients of ${\hat H}_{\rm boson}$ as input (the error induced by doing this can be suppressed~\cite{supmat}) to reshape the total Hamiltonian as a displacement channel. Homodyne measurement on the momentum quadrature of the resulting coherent state creates a quantum signal that grows linearly over time with a fixed variance. This achieves the Heisenberg limit. Implementing the robust frequency estimation (RFE)\cite{Haoya2023} on the quantum signal reduces the number of measurements to $\mathcal{O}({\rm polylog(\epsilon^{-1}) )}$, while making the algorithm robust under small SPAM error. We further develop an alternative scheme based on distributed quantum sensing (DQS) to ${\hat H}_{\rm int}$ that also achieves the Heisenberg limit, but with lower maximum evolution time. The general scheme of our approach is provided in Fig.~\ref{flowchart} with a detailed pseudo-code in the Supplementary Materials~\cite{supmat}.

\textit{Learning pure spin and boson coefficients}--- We begin by introducing the learning protocols for terms that only include pure spin or boson operators, which are characterized by  $\xi_a$ and $\omega_n$. A key algorithm we used is RUT, which reshape ${\hat H}$ by inserting random unitary sequence $\mathbb{U}(\boldsymbol \theta) = \prod_j U_j(\theta_j)$ between $R$ segments of $e^{-i{\hat H}\tau}$ ($\tau \equiv T/R$, and $\boldsymbol\theta \equiv (\theta_1, \theta_2, ..., \theta_j)$.). By sampling $\theta_j$ from independent uniform distribution $\mathcal{U}_j$ for each insertion, the effective Hamiltonian can be approximated as: $\hat{\mathcal H} = V({\boldsymbol \theta})^{-1}\int d{\boldsymbol\theta} \mathbb{U}^\dagger {\hat H} \mathbb{U}$, where $V({\boldsymbol \theta})^{-1}$ is the volumn of sampling domain. The form of $\hat{\mathcal H}$ depends on the choice of $\mathbb{U}$ used, which allows us to design the reshaping process. Readers are referred to~\cite{supmat} for technical details regarding RUT.

To cancel ${\hat H}_{\rm int}$, we reshape ${\hat H}$ with $\mathbb{U}^{(1)} \equiv \prod_n^{N_b} e^{-i \theta {\hat b}_n^\dagger {\hat b}_n}$ ($\theta \sim \mathcal{U}(0, 2\pi)$, where $N_b$ is the total number of bosonic mode. The application of $\mathbb{U}^{(1)}$ introduces $e^{ i\theta}$ phase to ${\hat b}^\dagger_n$ and $e^{ -i\theta}$ phase to ${\hat b}_n$, leading to the following effective Hamiltonian upon effective integration in RUT:
\begin{eqnarray}\label{Eff_H1}
\hat{\mathcal H}^{(1)} = \sum_a \xi_a \hat{E}_a +\sum_n \omega_n {\hat b}_n^\dagger {\hat b}_n
\end{eqnarray}
where the spin and bosonic DOFs are decoupled. This allows us to learn $\xi_a$ and $\omega_n$ separately.

(i.) To learn $\xi_a$, we begin by considering all Pauli string ${\hat E}_{b}$ such that $|{\rm supp}({\hat E}_{b})| = k$. For a general $k-$local spin Hamiltonian, the number of such ${\hat E}_{b}$ is $3^k{{N_q}\choose{k}}$ ($N_q$ is the number of qubit), which remains manageable assuming $k \sim \mathcal{O}(1)$. For each ${\hat E}_{b}$, we construct a unitary sequence $\mathbb{U}_b^{(2)} = \prod_{j}^{N_q} U_j$ , where the single-qubit unitary $U_j$ is defined as:
\begin{eqnarray}\label{pauli random}
U_{j}= 
    \begin{cases}
        e^{-i\theta_{j}\mathcal{P}_j^{b}} & \text{if } j \in {\rm supp}({\hat E}_{b})\\
        e^{-i\theta_{j}\mathcal{P}_j }e^{-i\phi_j\mathcal{P}_j^\prime} & \text{if } j \not\in {\rm supp}({\hat E}_{b})
    \end{cases}
\end{eqnarray}
Here,$\theta_j$ and $\phi_j$ are independently sampled from  uniform distribution $\mathcal{U}_j(0, \pi)$. $\mathcal{P}_j^{b}$ denotes the $j$-th Pauli operator in $E_{b}$. For example, if ${\hat E}_{b} = ZIX$, then$\mathcal{P}_{j=1}^{b} = Z$, $\mathcal{P}_{j=2}^{b} = I$ and $\mathcal{P}_{j=3}^{b} = X$. $\mathcal{P}_j$ and $\mathcal{P}_j^{\prime}$ are arbitrary Pauli operators satisfying $[\mathcal{P}_j, \mathcal{P}_j^{\prime}] \neq 0$. Reshaping $\hat{\mathcal H}^{(1)}$ with $\mathbb{U}_b^{(2)}$ gives the following effective spin Hamiltonian:
\begin{eqnarray}\label{Eff_Spin_H1}
\hat{\mathcal H}_{\rm S}^{(2)} &=& \sum_{s: {\hat E}_s \in S_{b}} \xi_s {\hat E}_s
\end{eqnarray}
where ${\hat E}_s$ is a Pauli string that belongs to the set $S_{b} \equiv \left\{\prod_{ i \in \mathrm{supp}({\hat E}_{b})} \mathcal{P}_i^s \;\middle|\; \mathcal{P}_i^s \in \{ \mathcal{P}_i^{b}, \mathcal{I} \} \right\}$. $\xi_s$ is the corresponding coefficient of ${\hat E}_s$ in $\hat{\mathcal H}^{(1)}$. Each ${\hat E}s$ is generated any subset of Pauli operators in ${\hat E}{b}$ by identity. Therefore, all elements in $S_{b}$ mutually commute. For instance, if ${\hat E}{b} = ZIX$, then $S_b = \{ZIX, ZII, IIX, III\}$. This reshaping exploits the commutation properties of Pauli operators (e.g. $e^{iX\theta}Ye^{-iX\theta} = {\rm cos}(2\theta)Y - {\rm sin}(2\theta)Z$), causing any Pauli string not in $S_{b}$ to acquire $\cos$ and $\sin$ coefficients that average to zero under RUT.

Since $|{\rm supp}({\hat E}_{b})| = k$ by selection, the number of terms in $\hat{\mathcal H}_{\rm S}^{(2)}$ is $|S_{b}| = 2^k$ . All these terms share the same eigenstates as ${\hat E}_{b}$. We label the eigenstate of ${\hat E}_{b}$ with $\ket{E_{b}}_l$, where the index $l$ ranges from $1$ to $2^k$ (as each Pauli operator has two eigenstates). For any ${\hat E}_s \in S_b$, we denote its eigenvalue on $\ket{E_b}_l$ as $\gamma_l^s$, such that ${\hat E}_s\ket{E_{b}}_l = \gamma_l^s \ket{E_{b}}_l$. $\gamma_l^s$ can only take the value of $\pm 1$. For example, if $E_{b} = ZIX$, let $\ket{E_{b}}_l = \ket{0}_1 \otimes \ket{-}_3$, the corresponding $\gamma_l^s$ for elements in $S_{b}$, as in previous example, are $\left\{-1, +1, -1, +1\right\}$ in order. As $\hat{\mathcal H}_{\rm S}^{(2)}$ sums up all ${\hat E}_s \in S_b$, the eigenvalue of $\ket{E_{b}}_l$ with respect to $\hat{\mathcal H}_{\rm S}^{(2)}$ can be written as: $\Xi_{b,l} = \sum_s \gamma_l^s \xi_s$. This allows us to implement the robust phase estimation (RPE)~\cite{Shelby2015} to learn $\Xi_{b,l}$ at Heisenberg limit. As $\ket{{\hat E}_{b}}_l$ and $\gamma_l^s$ are known by selecting a specific $E_b$, looping over all possible $l$ yields $2^{k}$ linear equations (LE). Solving these LEs simultaneously generates all $\xi_s$ such that ${\hat E}_s \in S_{b}$. This procedure should be repeated for all ${\hat E}_{b}$ that satisfies $|{\rm supp}({\hat E}_{b})| = k$.   Note that the above procedures can be parallelized on different devices, as all ${\hat E}_{b}$ and $\ket{{\hat E}_{b}}_l$ can be determined knowing $k$ and $N_q$. 

(ii) The learning of $\omega_{n}$ is rather straightforward, as the bosonic part of $\hat{\mathcal H}^{(1)}$ is a free-field Hamiltonian. We initialize the bosonic state on a coherent state with displacement $\alpha = |\alpha| e^{ir}$ and let it evolve under the bosonic part of $\hat{\mathcal H}^{(1)}$. The homodyne measurement on both the displacement and momentum quadrature of the $n^{th}$ bosonic mode allows us to construct a complex signal $\mathcal{Z} = \langle {\hat X}_n\rangle + i \langle {\hat P}_n\rangle$ with ${\rm arg}(\mathcal{Z}) = (\omega_{n} +r)t$ (as long as $\alpha$ is non-zero). Here, ${\hat P}_n \equiv i({\hat b}_n^\dagger - {\hat b}_n)/\sqrt{2}$ is the momentum quadrature of $n^{th}$ bosonic mode. This allows us to implement robust frequency estimation (RFE)\cite{Haoya2023}, which learns $\omega_{n}$ at Heisenberg limit. Note that the above process can be performed to all bosonic simultaneously by preparing a multi-mode coherent state.

\textit{Learning spin-boson couplings}.--- Now we proceed to the learning protocols for terms that contain a mixture of spin and boson operators, which are characterized by $\lambda_a^n$. We assume estimations ${\tilde\omega}_n$ have been obtained with small error from previous section. Starting from Eq.~\ref{gen_H}, to simplify the complex spin-boson interaction, we wish to reshape the spin-part of ${\hat H}_{\rm int}$ into a Hamiltonian with known eigenstates. This can be achieved by reshaping ${\hat H}$ with the same unitary sequence $\mathbb{U}_b^{(2)}$ ($\mathbb{U}^{(1)}$ is not applied to keep the spin-boson coupling terms), since the spin-part of ${\hat H}_{\rm int}$ shares the same interaction structure as ${\hat H}_{\rm spin}$. The resulting effective Hamiltonian is given by:
\begin{eqnarray}\label{Eff_H2}
\hat{\mathcal H}^{(2)}_{\rm SB} = \sum_{s: {\hat E}_s \in S_{b}} [\xi_s + \sum_n^{N_b} \lambda_s^n({\hat b}_{n}^\dagger + {\hat b}_{n}) ] {\hat E}_s + \sum_n^{N_b}\omega_n {\hat b}_n^\dagger {\hat b}_n \nonumber \\
\end{eqnarray}
where all ${\hat E}_s$ can be diagonalized simultaneously by $\ket{{\hat E}_{b}}_l$ ($l$ represents one of the $2^k$ eigenstates of ${\hat E}_{b}$). By initializing the spin-part of the wavefunction on $\ket{E_{b}}_l$, ${\hat E}_s$ in the above Hamiltonian can be replaced with their corresponding eigenvalues, leaving an effective bosonic Hamiltonian in the form of a series of displaced harmonic oscillators. This allows us to derive an analytical formula for the time evolution for bosonic states~\cite{supmat}:
\begin{eqnarray}\label{displacement}
\ket{\Psi_{\rm B}(t)}_l =  \prod_n^{N_b}{\hat D}_{n}(\frac{\Lambda_{b,l}^n}{\omega_{n}}(e^{-i\omega_{n} t} - 1)) \ket{\Psi_{\rm B}(0)} 
\end{eqnarray}
where $\Lambda_{b,l}^n \equiv \sum_s \gamma_{l}^s \lambda_{s}^n$ is the eigenvalue of $\ket{E_{b}}_l$ with respect to the spin-part of ${\hat H}_{\rm int}$. $\ket{\Psi_{\rm B}(0)}$ is the initial bosonic state, which can take the form of coherent state or squeezed coherent state. ${\hat D}_{n}(\alpha) \equiv e^{\alpha {\hat b}_{n}^\dagger - \alpha^\ast {\hat b}_{n}}$ is the displacement operator of the $n^{th}$ bosonic mode. Starting from Eq.~\ref{displacement}, we provide two solutions to learn $\lambda_{s}^n$ at Heisenberg limit, based on Trotterization technique and distributed quantum sensing method respectively:
\begin{figure}[t]
\centering
\centerline{\includegraphics[width=100mm]{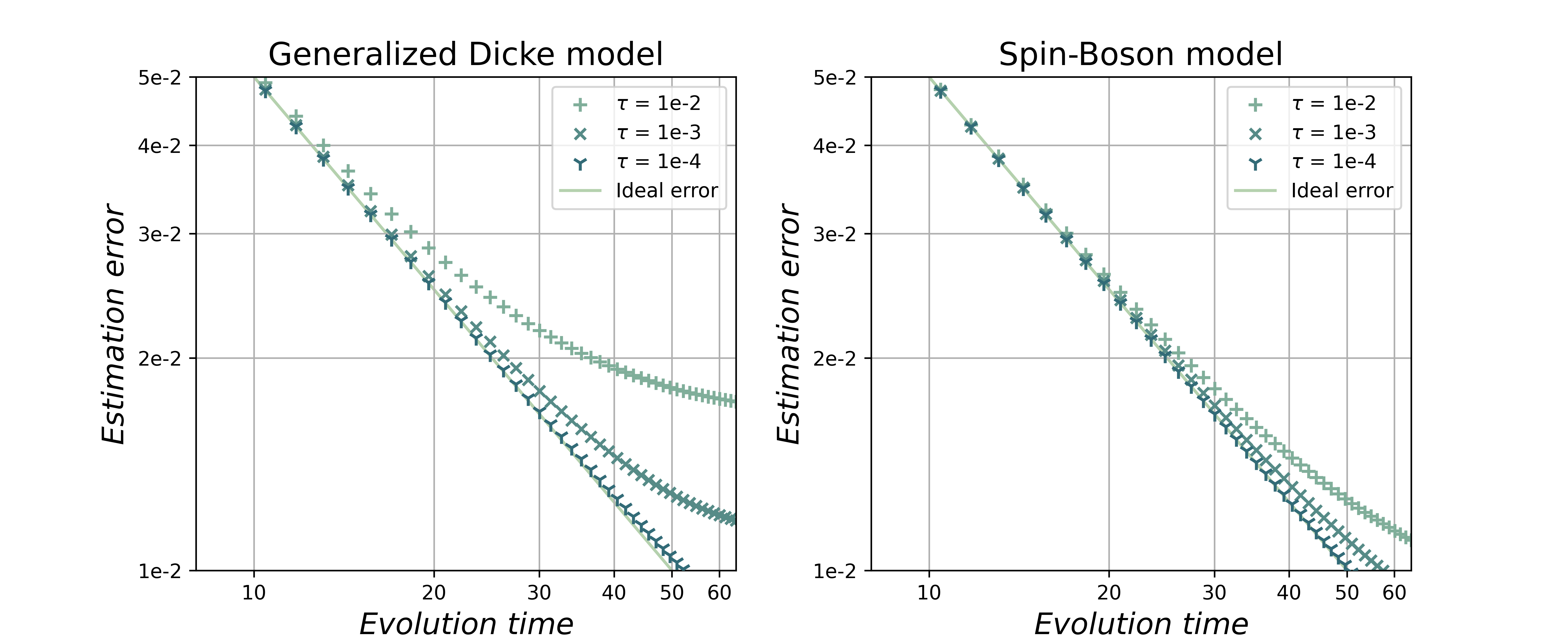}}
\caption{Estimation error scaling at different $\tau$. The presented data is averaged over $100$ independent runs. Ideal error refers to $\epsilon = 1/2T$, which is derived from Eq.~\ref{displacement}. For GDM, the target parameter is $\lambda_{XXI}$. For SBM, the target parameter is $\lambda_X^{n=1}$.}
\label{Trotter_error}
\end{figure}

(i): Trotter-based scheme. The trotter-based scheme cancel out $\omega_{n}$ via trotterization. We construct a unitary sequence $\mathbb{U}^{(3)}= \sum_n e^{i {\tilde \omega}_{n} {\hat b}_{n}^\dagger  {\hat b}_{n} \tau}$ using the estimated ${\tilde \omega}_{n}$ from the previous section. $\mathbb{U}^{(3)}$ is inserted at the end of every repetitive unitary cycles in RUT. By choosing $\ket{\Psi_{\rm B}(0)} = \ket{\boldsymbol 0}$ (bosonic vacuum state), the actual state after $R$ cycles can be written as:
\begin{eqnarray}\label{insert}
\ket{\Psi_{\rm SB}(t)}_{l} = \prod_i^R \Big[[\mathbb{U}^{(2)}_b({\boldsymbol \theta_i})]^\dagger e^{-i{\hat H}\tau}\mathbb{U}^{(2)}_b({\boldsymbol \theta_i})\mathbb{U}^{(3)} \ket{{\hat E}_b}_l\ket{\boldsymbol 0} \nonumber\\ 
\end{eqnarray}
where ${\boldsymbol \theta_i}$ represents all $\theta$ that parametrize $\mathbb{U}^{(2)}$. Note that $\mathbb{U}^{(3)}$ is not $\theta$ dependent. In the limit $R \rightarrow \infty$, the above equation effectively cancels the term $\sum_n \omega_{n} {\hat b}_{n}^\dagger  {\hat b}_{n}$ in $\hat{\mathcal H}^{(2)}_{\rm SB}$. The resulting bosonic wavefunction is: $\ket{\Psi_{\rm B}^{\rm eff}(t)}_l = \prod_n^{N_b}{\hat D}_{n}(-i\Lambda_{b,l}^n t) \ket{\bf 0}$ . The homodyne measurement on the momentum quadrature of bosonic modes gives $\langle \hat P_n \rangle = -\sqrt{2}\Lambda_{b,l}^n t$ with ${\rm Var}[ \hat P_n ] = 1$. This result follows from the fact that a coherent state always satisfies the minimal uncertainty, which achieves the Heisenberg limit in learning $\Lambda_{b,l}^n$. Looping over all possible $l$ generates all $\lambda_s^n$ with ${\hat E}_s \in S_b$, same as the learning protocol for $\xi_s$. One might be tempted to use $\langle \hat P_n \rangle/\sqrt{2}t$ as the estimator of $\Lambda_{b,l}^n$ . However, doing so would lead to a $\mathcal{O}(\epsilon^{-2})$ scaling in the number of measurements. To improve this, we construct a signal $\mathcal{Z} = e^{-i \langle \hat P_n \rangle}$ via post-processing\cite{supmat}, which allows us to implement RFE\cite{Haoya2023}, achieving a $\mathcal{O}({\rm polylog(\epsilon^{-1}) )}$ scaling with respect to the number of measurements.

\begin{figure}[t]
\centering
\centerline{\includegraphics[width=100mm]{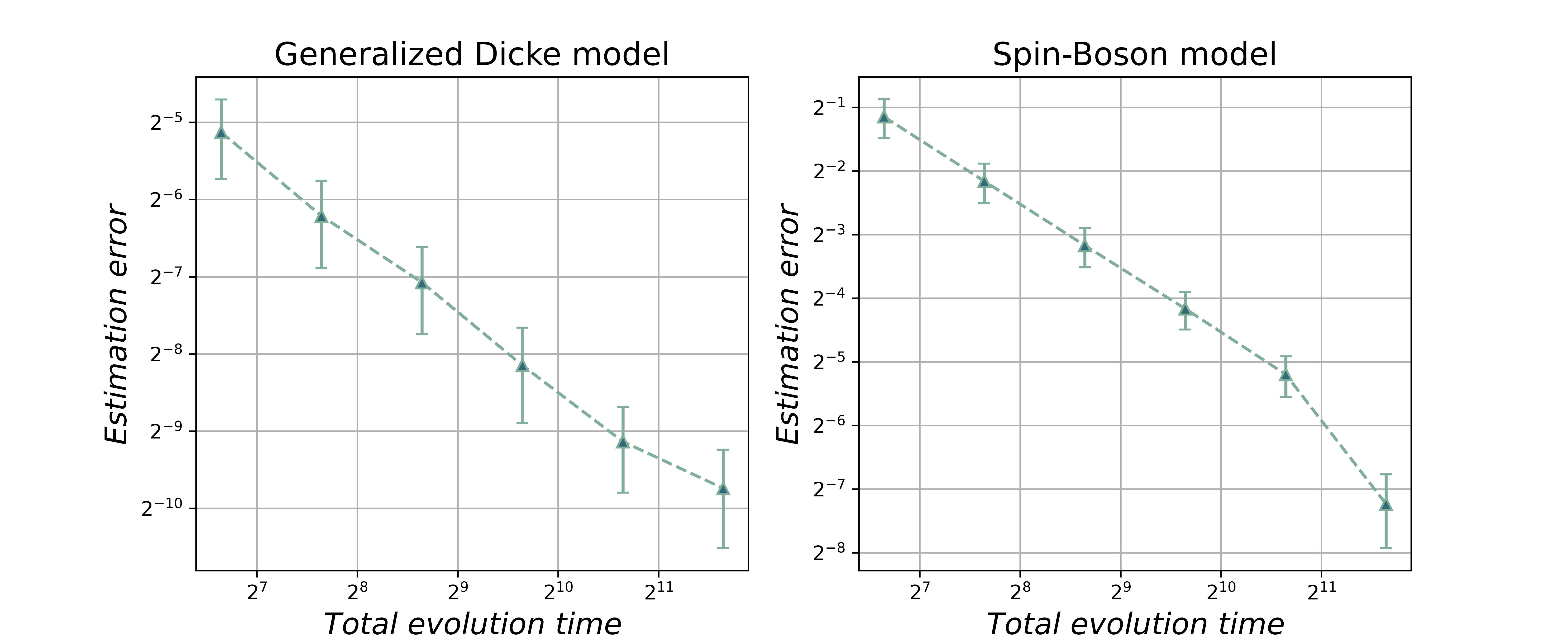}}
\caption{Estimation error scaling with SPAM error. For GDM, the target parameter is $\lambda_{XXI}$. For SBM, the target parameter is $\lambda_X^{n=1}$. $\mathfrak K = 6$ is used in the $2^{\mathfrak K}$ fold space to implement RFE.}
\label{RPE_error}
\end{figure}

(ii): Distributed quantum sensing scheme. Starting from Eq.~\ref{displacement}, the Heisenberg limit can also be reached by initializing the wavefunction on an entangled bosonic state. By preparing $W$ copies of the unknown Hamiltonian, an entangled squeezed state can be prepared via a balanced beam splitter\cite{Quntao2018}:
\begin{eqnarray}\label{multimode_entangled}
\ket{\Psi_{\rm B}^{\rm ent}} &=& \prod_n^{N_b}{\rm exp}[\frac{1}{2}(z^\ast {\hat B}_{n}^2 - z {\hat B}_{n}^{\dagger 2})] \ket{\boldsymbol 0}
\end{eqnarray}
where ${\hat B}_n \equiv \sum_{w=1}^{W} {\hat b_{n, w}} / \sqrt{W}$ is the entangled annihilation operator of bosonic mode $n$.  As the bosonic part of $\hat{\mathcal H}^{(2)}_{\rm SB}$ contains no non-linear terms, $e^{-i\hat{\mathcal H}^{(2)}_{\rm SB}t}$ can be treated as a pure displacement channel if we choose $\ket{\Psi_{\rm B}(0)} = \ket{\Psi_{\rm B}^{\rm ent}}$ in Eq.~\ref{displacement}~\cite{supmat}. Let ${\tilde{\mathcal X}}_n \equiv \sum_{w=1}^W \langle{\hat X_{n, w}}\rangle/W$ be the displacement estimator, at $t = \pi/{\tilde \omega}_{n}$, we have $\mathbb{E}[{\tilde{\mathcal X}}_n] = -2\sqrt{2} \Lambda_{b,l}^n/\omega_{n}$, and the RMSE of ${\tilde{\mathcal X}}_n$ is\cite{Quntao2018}:
\begin{eqnarray}\label{DQS}
\epsilon({\tilde{\mathcal X}}_n) = \sqrt{\Big(\frac{1}{4W(\sqrt{1+N_{\rm pt}} + \sqrt{N_{\rm pt}})^2}\Big)}
\end{eqnarray}
where $N_{\rm pt} = {\rm sinh}^2(|z|)$ is the total photon number. The sum of evolution time across all copies is $T = W\pi/{\tilde \omega_{n}}$. If the mean photon number per node $n_{\rm pt} \equiv N_{\rm pt}/W$ is fixed, we have $\epsilon({\tilde{\mathcal X}}_n) \sim \mathcal{O}(W^{-1})$. Therefore, we have $\epsilon({\tilde{\mathcal X}}_n) \sim \mathcal{O}(T^{-1})$, which reaches the Heisenberg limit. 

Comparing the above two schemes, the former requires only $\mathcal{O}({\rm polylog(\epsilon^{-1}) )}$ number of measurements, in contrast to the $\mathcal{O}(\epsilon^{-1})$ scaling of the latter. However, the $\mathcal{O}({\rm polylog(\epsilon^{-1}) )}$ scaling of the former relies on the implementation of RFE, which requires time sampling over
$2^{\mathfrak{K}}-$fold space. This necessitates a significantly higher number of gate applied to suppress errors induced during RUT ,while the latter require only a maximum evolution time of $\pi/{\tilde \omega}_{n}$. Nevertheless, both schemes achieve the Heisenberg limit. 

\textit{Numerical experiments}.--- Our protocol can be applied to learn a wide range of models from AMO physics to condensed matter (e.g. electron-phonon interaction) and high energy physics (e.g. electron-photon interaction). Some examples of learnable models are listed in  Table~\ref{tab:models}. Here, we demonstrate the application of our algorithm on two specific classes of models that are important in AMO physics, which can be realized in cavity QED and circuit QED: a generalized Dicke model (GDM) and a spin-boson model (SBM). 
\begin{table*}[t]
\centering
\begin{tabular}{||c | c | c | c||} 
 \hline
 Model Name & ${\hat H}_{\rm spin}$ & ${\hat H}_{\rm boson}$ & ${\hat H}_{\rm int}$ \\ [0.5ex] 
 \hline\hline
 1D Holstein & $\sum_{\langle i,j \rangle}\xi_{i,j}(X_i X_j + Y_i Y_j)$ & $\sum_i \lambda_i (1-Z_i)({\hat b}_i^\dagger + {\hat b}_i)$ & $\sum_i \omega_i {\hat b}_i^\dagger {\hat b}_i$ \\ 
 \hline
 1D SSH & $\sum_i \xi_{i}(1-Z_i)$ & $ \sum_{\langle i,j \rangle} \lambda_{i,j} (X_i X_j + Y_i Y_j)({\hat b}_{i,j}^\dagger + {\hat b}_{i,j})$ & $\sum_{\langle i,j \rangle} \omega_{i,j} {\hat b}_{i,j}^\dagger {\hat b}_{i,j}$ \\ 
 \hline
 Spin-Peierls & $\sum_{i}\xi_i \vec{S}_i \vec{S}_{i+1}$ & $\sum_{i}\lambda_i({\hat X}_{i+1} -{\hat X}_i)\vec{S}_i \vec{S}_{i+1}$ & $\sum_i \omega_i {\hat b}_i^\dagger {\hat b}_i$ \\ 
 \hline
\end{tabular}
\caption{Models learnable by our protocol\cite{Holstein_model, SSH_model, SPmodel}. The 1D Holstein and SSH electron-phonon Hamiltonians are obtained via Jordan-Wigner transformation, where $\vec{S}$ denotes the spin-$1/2$ operator related to Pauli matrices up to coefficients. The Schwinger model\cite{Schwinger_model} for quantum electrodynamics can also be incorporated upon minor changes to our protocol\cite{supmat}.}
\label{tab:models}
\end{table*}
The Hamiltonian of GDM is given by:
\begin{eqnarray}
{\hat H}_{\rm GDM} =  \sum_{a}[ \xi_a + \lambda_a ({\hat b}^\dagger + {\hat b}) ]{\hat E}_a + \omega {\hat b}^\dagger{\hat b}
\end{eqnarray}
where ${\hat E}_a \in \Big\{\mathcal{P}\mathcal{P}I, \mathcal{P}I\mathcal{P}, I\mathcal{P}\mathcal{P} | \mathcal{P} \in \{X,Y,Z\}\Big\}$ is a 3-qubit Pauli string. We set $\omega = 1$, $\xi_{a}$  and $\lambda_a$ are uniformly sampled from $\mathcal{U}(0.5, 1.5)$ and $\mathcal{U}(0.01, 0.03)$, respectively. The GDM describes the coupling between a inhomogeneous Heisenberg spin chain and a single bosonic mode, which provides the theoretical description for a range of quantum devices and algorithms~\cite{liu2024hybrid, safavi2018verification}.

For SBM, we have:
\begin{eqnarray}\label{SBM}
{\hat H}_{\rm SBM} =  \sum_{a}[ \xi_a + \sum_n^{N_b} \lambda_a^n ({\hat b}_n^\dagger + {\hat b}_n) ]E_a + \sum_n^{N_b}\omega_n {\hat b}_n^\dagger{\hat b}_n \nonumber \\
\end{eqnarray}
where ${\hat E}_a \in \{X, Y, Z\}$ is a 1-qubit Pauli operator. $\xi_a \sim \mathcal{U}(0.5, 1.5)$. $\lambda_a^n = \kappa_a \Lambda_n$ with $\kappa_a$ sampled from $\mathcal{U}(0.5, 1.5)$. $\Lambda_n$ and $\omega_n$ are generated by discretizing Eq.~\ref{Jw}. SBM describes a non-markovian dissipation of a single qubit mediated via a bosonic bath, which is crucial for the simulation of decoherence~\cite{magazzu2018probing} and quantum phase transition~\cite{de2020quantum}.

\begin{figure}[b]
\centerline{\includegraphics[width=80mm]{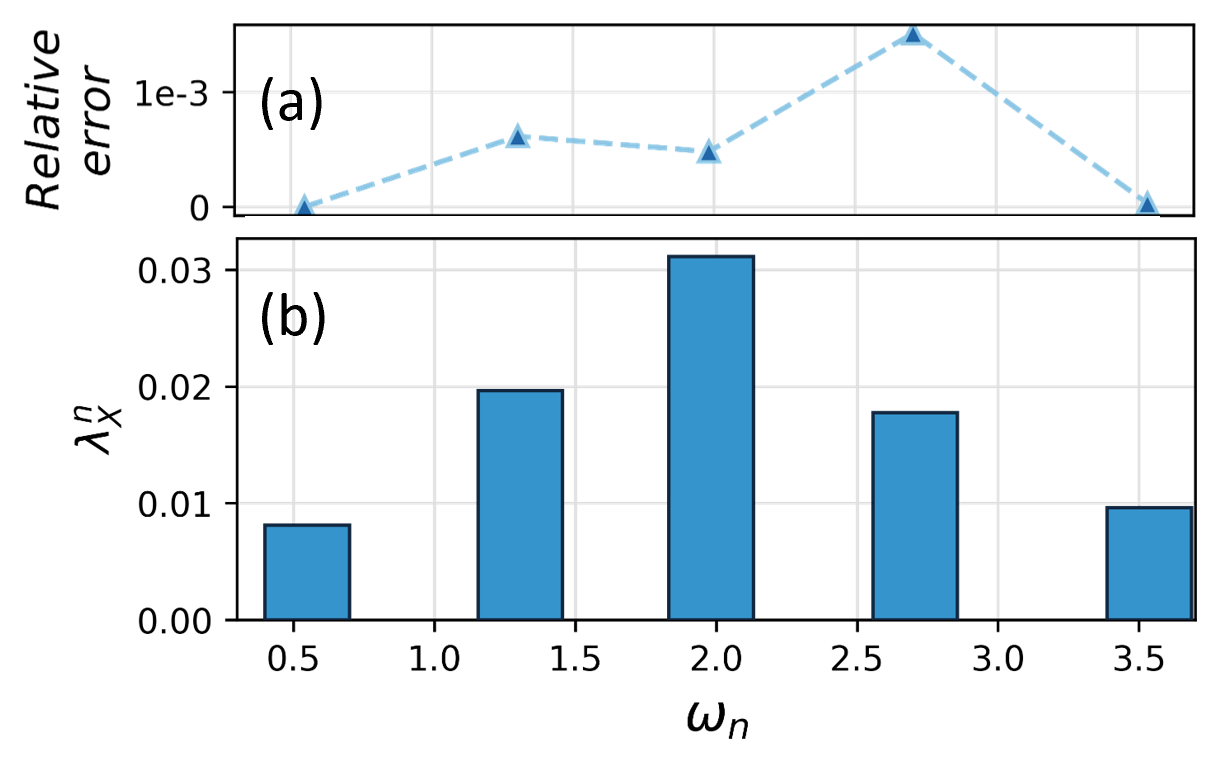}}
\caption{Learning the spectral density function of SBM. (a) Relative error of the estimation generated from the DQS-based scheme. (b) The value of $\lambda_X^{n}$ across all bosonic modes, showing the shape of the discrete spectral density $J(\omega)$. For the parameter used: $W = 1000$ and $n_{pt} = 1$.}
\label{Spectral_learning}
\end{figure}

In Fig.~\ref{Trotter_error}, to demonstrate the error scaling induced by RUT, we plot the total evolution time against mean estimation error, which is obtained by averaging 100 numerical experiments. The time step $\tau =  t/R$ is changed across different lines. We observe that as $\tau$ decreases, the error induced by RUT gradually aligns with the ideal error (gray line). Further numerical experiment shows that the trace distance between the ideal and actual state follows $||\Delta \rho ||_1 \propto t/\sqrt{R}$~\cite{supmat}. 

In Fig.~\ref{RPE_error}, we demonstrate the robustness of our algorithm against SPAM error using the trotter-based scheme. A small Gaussian noise is added into measurements, and $6$ time points are sampled in $2^\mathfrak{K}-$fold space to implement RFE. We observe that the estimation error maintains a Heisenberg scaling despite the presence of noise.

In Fig.~\ref{Spectral_learning}, we demonstrate the learning of unknown bosonic spectrum. The spectral density function, describing a cavity mode coupled to a dissipative bosonic heat bath~\cite{Spec_den}, takes the form of:
\begin{eqnarray}\label{Jw}
J(\omega) = \sum_n (\Lambda_n)^2 \delta(\omega-\omega_n) = \frac{\eta \omega}{(\omega^2-\Omega^2)^2 + \gamma^2 \omega^2} \nonumber \\
\end{eqnarray}
where $\eta = 0.01$, $\gamma = 1$ and $\Omega = 2$. $\Lambda_n$ and $\omega_n$ are generated by discretizing $J(\omega)$~\cite{supmat}. Using the DQS-based scheme, we reconstruct the spectral density function by looping over all discrete bosonic modes. The simulation is performed with $W = 1000$ and $n_{pt} = 1$. The spectral density is recovered within $0.2\%$ error, showcasing the applicability of our algorithm in spectrum learning.

\textit{Discussion}.---In this work, we establish a robust and efficient foundation for hybrid Hamiltonian learning. Our protocol reaches the golden-standard: Heisenberg limit $T \sim \mathcal{O}(\epsilon^{-1})$, in learning all Hamiltonian coefficients. It only requires $\mathcal{O}({\rm polylog}(\epsilon^{-1}))$ measurements while remaining robust under small SPAM error. In learning pure spin and boson coefficients, our algorithm operates without relying on the low-intersection approximation. In learning spin-boson coupling coefficients, we propose two schemes: trotter-based scheme and DQS-based scheme. The former employs RFE, achieving the Heisenberg limit using only $\mathcal{O}({\rm polylog}(\epsilon^{-1}))$. The latter utilizes quantum entanglement, requiring a significantly shorter maximum evolution time, making it particularly well-suited for detecting transient couplings. Our work can be widely adapted for a variety of models from AMO physics to condensed matter and high energy physics. For future exploration, one can consider generalizing our learning protocol to high-dimensional fermionic Hamiltonian and non-linear bosonic modes, which will be significant for mixed-species quantum system learning. It also opens up a number of exciting directions for enhancing quantum simulations with quantum learning theory in hybrid quantum systems~\cite{peng2024provably}, where our learning algorithms could provide precise real-time information in quantum sensing and feedback control.

\textit{Acknowledgement}.---The authors thank Yu Tong and Yukai Wang for valuable discussion. This work used computational and storage services associated with the Hoffman2 Cluster which is operated by the UCLA Office of Advanced Research Computing’s Research Technology Group.

\bibliography{main}

\begin{thebibliography}{34}%
\makeatletter
\providecommand \@ifxundefined [1]{%
 \@ifx{#1\undefined}
}%
\providecommand \@ifnum [1]{%
 \ifnum #1\expandafter \@firstoftwo
 \else \expandafter \@secondoftwo
 \fi
}%
\providecommand \@ifx [1]{%
 \ifx #1\expandafter \@firstoftwo
 \else \expandafter \@secondoftwo
 \fi
}%
\providecommand \natexlab [1]{#1}%
\providecommand \enquote  [1]{``#1''}%
\providecommand \bibnamefont  [1]{#1}%
\providecommand \bibfnamefont [1]{#1}%
\providecommand \citenamefont [1]{#1}%
\providecommand \href@noop [0]{\@secondoftwo}%
\providecommand \href [0]{\begingroup \@sanitize@url \@href}%
\providecommand \@href[1]{\@@startlink{#1}\@@href}%
\providecommand \@@href[1]{\endgroup#1\@@endlink}%
\providecommand \@sanitize@url [0]{\catcode `\\12\catcode `\$12\catcode `\&12\catcode `\#12\catcode `\^12\catcode `\_12\catcode `\%12\relax}%
\providecommand \@@startlink[1]{}%
\providecommand \@@endlink[0]{}%
\providecommand \url  [0]{\begingroup\@sanitize@url \@url }%
\providecommand \@url [1]{\endgroup\@href {#1}{\urlprefix }}%
\providecommand \urlprefix  [0]{URL }%
\providecommand \Eprint [0]{\href }%
\providecommand \doibase [0]{http://dx.doi.org/}%
\providecommand \selectlanguage [0]{\@gobble}%
\providecommand \bibinfo  [0]{\@secondoftwo}%
\providecommand \bibfield  [0]{\@secondoftwo}%
\providecommand \translation [1]{[#1]}%
\providecommand \BibitemOpen [0]{}%
\providecommand \bibitemStop [0]{}%
\providecommand \bibitemNoStop [0]{.\EOS\space}%
\providecommand \EOS [0]{\spacefactor3000\relax}%
\providecommand \BibitemShut  [1]{\csname bibitem#1\endcsname}%
\let\auto@bib@innerbib\@empty
\bibitem [{\citenamefont {Bardeen}\ \emph {et~al.}(1957)\citenamefont {Bardeen}, \citenamefont {Cooper},\ and\ \citenamefont {Schrieffer}}]{bardeen1957theory}%
  \BibitemOpen
  \bibfield  {author} {\bibinfo {author} {\bibfnamefont {J.}~\bibnamefont {Bardeen}}, \bibinfo {author} {\bibfnamefont {L.~N.}\ \bibnamefont {Cooper}}, \ and\ \bibinfo {author} {\bibfnamefont {J.~R.}\ \bibnamefont {Schrieffer}},\ }\href@noop {} {\bibfield  {journal} {\bibinfo  {journal} {Physical review}\ }\textbf {\bibinfo {volume} {108}},\ \bibinfo {pages} {1175} (\bibinfo {year} {1957})}\BibitemShut {NoStop}%
\bibitem [{\citenamefont {Feynman}(2018)}]{feynman2018space}%
  \BibitemOpen
  \bibfield  {author} {\bibinfo {author} {\bibfnamefont {R.~P.}\ \bibnamefont {Feynman}},\ }in\ \href@noop {} {\emph {\bibinfo {booktitle} {Quantum Electrodynamics}}}\ (\bibinfo  {publisher} {CRC Press},\ \bibinfo {year} {2018})\ pp.\ \bibinfo {pages} {178--198}\BibitemShut {NoStop}%
\bibitem [{\citenamefont {Schwinger}(1951)}]{schwinger1951theory}%
  \BibitemOpen
  \bibfield  {author} {\bibinfo {author} {\bibfnamefont {J.}~\bibnamefont {Schwinger}},\ }\href@noop {} {\bibfield  {journal} {\bibinfo  {journal} {Physical Review}\ }\textbf {\bibinfo {volume} {82}},\ \bibinfo {pages} {914} (\bibinfo {year} {1951})}\BibitemShut {NoStop}%
\bibitem [{\citenamefont {Yang}\ and\ \citenamefont {Mills}(1954)}]{yang1954conservation}%
  \BibitemOpen
  \bibfield  {author} {\bibinfo {author} {\bibfnamefont {C.-N.}\ \bibnamefont {Yang}}\ and\ \bibinfo {author} {\bibfnamefont {R.~L.}\ \bibnamefont {Mills}},\ }\href@noop {} {\bibfield  {journal} {\bibinfo  {journal} {Physical review}\ }\textbf {\bibinfo {volume} {96}},\ \bibinfo {pages} {191} (\bibinfo {year} {1954})}\BibitemShut {NoStop}%
\bibitem [{\citenamefont {Enkner}\ \emph {et~al.}(2024)\citenamefont {Enkner}, \citenamefont {Graziotto}, \citenamefont {Bori{\c{c}}i}, \citenamefont {Appugliese}, \citenamefont {Reichl}, \citenamefont {Scalari}, \citenamefont {Regnault}, \citenamefont {Wegscheider}, \citenamefont {Ciuti},\ and\ \citenamefont {Faist}}]{FQHE}%
  \BibitemOpen
  \bibfield  {author} {\bibinfo {author} {\bibfnamefont {J.}~\bibnamefont {Enkner}}, \bibinfo {author} {\bibfnamefont {L.}~\bibnamefont {Graziotto}}, \bibinfo {author} {\bibfnamefont {D.}~\bibnamefont {Bori{\c{c}}i}}, \bibinfo {author} {\bibfnamefont {F.}~\bibnamefont {Appugliese}}, \bibinfo {author} {\bibfnamefont {C.}~\bibnamefont {Reichl}}, \bibinfo {author} {\bibfnamefont {G.}~\bibnamefont {Scalari}}, \bibinfo {author} {\bibfnamefont {N.}~\bibnamefont {Regnault}}, \bibinfo {author} {\bibfnamefont {W.}~\bibnamefont {Wegscheider}}, \bibinfo {author} {\bibfnamefont {C.}~\bibnamefont {Ciuti}}, \ and\ \bibinfo {author} {\bibfnamefont {J.}~\bibnamefont {Faist}},\ }\href@noop {} {\bibfield  {journal} {\bibinfo  {journal} {arXiv preprint arXiv:2405.18362}\ } (\bibinfo {year} {2024})}\BibitemShut {NoStop}%
\bibitem [{\citenamefont {Song}\ \emph {et~al.}(2023)\citenamefont {Song}, \citenamefont {Goldman},\ and\ \citenamefont {Fu}}]{TI}%
  \BibitemOpen
  \bibfield  {author} {\bibinfo {author} {\bibfnamefont {X.-Y.}\ \bibnamefont {Song}}, \bibinfo {author} {\bibfnamefont {H.}~\bibnamefont {Goldman}}, \ and\ \bibinfo {author} {\bibfnamefont {L.}~\bibnamefont {Fu}},\ }\href@noop {} {\bibfield  {journal} {\bibinfo  {journal} {Physical Review B}\ }\textbf {\bibinfo {volume} {108}},\ \bibinfo {pages} {205123} (\bibinfo {year} {2023})}\BibitemShut {NoStop}%
\bibitem [{\citenamefont {Savary}\ and\ \citenamefont {Balents}(2016)}]{savary2016quantum}%
  \BibitemOpen
  \bibfield  {author} {\bibinfo {author} {\bibfnamefont {L.}~\bibnamefont {Savary}}\ and\ \bibinfo {author} {\bibfnamefont {L.}~\bibnamefont {Balents}},\ }\href@noop {} {\bibfield  {journal} {\bibinfo  {journal} {Reports on Progress in Physics}\ }\textbf {\bibinfo {volume} {80}},\ \bibinfo {pages} {016502} (\bibinfo {year} {2016})}\BibitemShut {NoStop}%
\bibitem [{\citenamefont {Xu}\ \emph {et~al.}(2024)\citenamefont {Xu}, \citenamefont {Li}, \citenamefont {Wang}, \citenamefont {Tang}, \citenamefont {Cappellaro},\ and\ \citenamefont {Li}}]{QT}%
  \BibitemOpen
  \bibfield  {author} {\bibinfo {author} {\bibfnamefont {H.}~\bibnamefont {Xu}}, \bibinfo {author} {\bibfnamefont {C.}~\bibnamefont {Li}}, \bibinfo {author} {\bibfnamefont {G.}~\bibnamefont {Wang}}, \bibinfo {author} {\bibfnamefont {H.}~\bibnamefont {Tang}}, \bibinfo {author} {\bibfnamefont {P.}~\bibnamefont {Cappellaro}}, \ and\ \bibinfo {author} {\bibfnamefont {J.}~\bibnamefont {Li}},\ }\href@noop {} {\bibfield  {journal} {\bibinfo  {journal} {Physical Review B}\ }\textbf {\bibinfo {volume} {110}},\ \bibinfo {pages} {085136} (\bibinfo {year} {2024})}\BibitemShut {NoStop}%
\bibitem [{\citenamefont {Stolk}\ \emph {et~al.}(2024)\citenamefont {Stolk}, \citenamefont {van~der Enden}, \citenamefont {Slater}, \citenamefont {te~Raa-Derckx}, \citenamefont {Botma}, \citenamefont {van Rantwijk}, \citenamefont {Biemond}, \citenamefont {Hagen}, \citenamefont {Herfst}, \citenamefont {Koek} \emph {et~al.}}]{Metro_scale}%
  \BibitemOpen
  \bibfield  {author} {\bibinfo {author} {\bibfnamefont {A.~J.}\ \bibnamefont {Stolk}}, \bibinfo {author} {\bibfnamefont {K.~L.}\ \bibnamefont {van~der Enden}}, \bibinfo {author} {\bibfnamefont {M.-C.}\ \bibnamefont {Slater}}, \bibinfo {author} {\bibfnamefont {I.}~\bibnamefont {te~Raa-Derckx}}, \bibinfo {author} {\bibfnamefont {P.}~\bibnamefont {Botma}}, \bibinfo {author} {\bibfnamefont {J.}~\bibnamefont {van Rantwijk}}, \bibinfo {author} {\bibfnamefont {J.~B.}\ \bibnamefont {Biemond}}, \bibinfo {author} {\bibfnamefont {R.~A.}\ \bibnamefont {Hagen}}, \bibinfo {author} {\bibfnamefont {R.~W.}\ \bibnamefont {Herfst}}, \bibinfo {author} {\bibfnamefont {W.~D.}\ \bibnamefont {Koek}},  \emph {et~al.},\ }\href@noop {} {\bibfield  {journal} {\bibinfo  {journal} {Science advances}\ }\textbf {\bibinfo {volume} {10}},\ \bibinfo {pages} {eadp6442} (\bibinfo {year} {2024})}\BibitemShut {NoStop}%
\bibitem [{\citenamefont {Konno}\ \emph {et~al.}(2024)\citenamefont {Konno}, \citenamefont {Asavanant}, \citenamefont {Hanamura}, \citenamefont {Nagayoshi}, \citenamefont {Fukui}, \citenamefont {Sakaguchi}, \citenamefont {Ide}, \citenamefont {China}, \citenamefont {Yabuno}, \citenamefont {Miki} \emph {et~al.}}]{GKP}%
  \BibitemOpen
  \bibfield  {author} {\bibinfo {author} {\bibfnamefont {S.}~\bibnamefont {Konno}}, \bibinfo {author} {\bibfnamefont {W.}~\bibnamefont {Asavanant}}, \bibinfo {author} {\bibfnamefont {F.}~\bibnamefont {Hanamura}}, \bibinfo {author} {\bibfnamefont {H.}~\bibnamefont {Nagayoshi}}, \bibinfo {author} {\bibfnamefont {K.}~\bibnamefont {Fukui}}, \bibinfo {author} {\bibfnamefont {A.}~\bibnamefont {Sakaguchi}}, \bibinfo {author} {\bibfnamefont {R.}~\bibnamefont {Ide}}, \bibinfo {author} {\bibfnamefont {F.}~\bibnamefont {China}}, \bibinfo {author} {\bibfnamefont {M.}~\bibnamefont {Yabuno}}, \bibinfo {author} {\bibfnamefont {S.}~\bibnamefont {Miki}},  \emph {et~al.},\ }\href@noop {} {\bibfield  {journal} {\bibinfo  {journal} {Science}\ }\textbf {\bibinfo {volume} {383}},\ \bibinfo {pages} {289} (\bibinfo {year} {2024})}\BibitemShut {NoStop}%
\bibitem [{\citenamefont {Lee}\ \emph {et~al.}(2024)\citenamefont {Lee}, \citenamefont {Kang}, \citenamefont {Lee}, \citenamefont {Jeong}, \citenamefont {Jiang},\ and\ \citenamefont {Lee}}]{HybridQC1}%
  \BibitemOpen
  \bibfield  {author} {\bibinfo {author} {\bibfnamefont {J.}~\bibnamefont {Lee}}, \bibinfo {author} {\bibfnamefont {N.}~\bibnamefont {Kang}}, \bibinfo {author} {\bibfnamefont {S.-H.}\ \bibnamefont {Lee}}, \bibinfo {author} {\bibfnamefont {H.}~\bibnamefont {Jeong}}, \bibinfo {author} {\bibfnamefont {L.}~\bibnamefont {Jiang}}, \ and\ \bibinfo {author} {\bibfnamefont {S.-W.}\ \bibnamefont {Lee}},\ }\href@noop {} {\bibfield  {journal} {\bibinfo  {journal} {PRX Quantum}\ }\textbf {\bibinfo {volume} {5}},\ \bibinfo {pages} {030322} (\bibinfo {year} {2024})}\BibitemShut {NoStop}%
\bibitem [{\citenamefont {Araz}\ \emph {et~al.}(2024)\citenamefont {Araz}, \citenamefont {Grau}, \citenamefont {Montgomery},\ and\ \citenamefont {Ringer}}]{HybridQC2}%
  \BibitemOpen
  \bibfield  {author} {\bibinfo {author} {\bibfnamefont {J.~Y.}\ \bibnamefont {Araz}}, \bibinfo {author} {\bibfnamefont {M.}~\bibnamefont {Grau}}, \bibinfo {author} {\bibfnamefont {J.}~\bibnamefont {Montgomery}}, \ and\ \bibinfo {author} {\bibfnamefont {F.}~\bibnamefont {Ringer}},\ }\href@noop {} {\bibfield  {journal} {\bibinfo  {journal} {arXiv preprint arXiv:2410.07346}\ } (\bibinfo {year} {2024})}\BibitemShut {NoStop}%
\bibitem [{\citenamefont {Santagati}\ \emph {et~al.}(2017)\citenamefont {Santagati}, \citenamefont {Wang}, \citenamefont {Paesani}, \citenamefont {Knauer}, \citenamefont {Gentile}, \citenamefont {Wiebe}, \citenamefont {Petruzzella}, \citenamefont {O’Brien}, \citenamefont {Rarity}, \citenamefont {Laing} \emph {et~al.}}]{HL1}%
  \BibitemOpen
  \bibfield  {author} {\bibinfo {author} {\bibfnamefont {R.}~\bibnamefont {Santagati}}, \bibinfo {author} {\bibfnamefont {J.}~\bibnamefont {Wang}}, \bibinfo {author} {\bibfnamefont {S.}~\bibnamefont {Paesani}}, \bibinfo {author} {\bibfnamefont {S.}~\bibnamefont {Knauer}}, \bibinfo {author} {\bibfnamefont {A.}~\bibnamefont {Gentile}}, \bibinfo {author} {\bibfnamefont {N.}~\bibnamefont {Wiebe}}, \bibinfo {author} {\bibfnamefont {M.}~\bibnamefont {Petruzzella}}, \bibinfo {author} {\bibfnamefont {J.}~\bibnamefont {O’Brien}}, \bibinfo {author} {\bibfnamefont {J.}~\bibnamefont {Rarity}}, \bibinfo {author} {\bibfnamefont {A.}~\bibnamefont {Laing}},  \emph {et~al.},\ }in\ \href@noop {} {\emph {\bibinfo {booktitle} {Frontiers in Optics}}}\ (\bibinfo {organization} {Optica Publishing Group},\ \bibinfo {year} {2017})\ pp.\ \bibinfo {pages} {FTh3E--7}\BibitemShut {NoStop}%
\bibitem [{\citenamefont {Guo}\ \emph {et~al.}(2025)\citenamefont {Guo}, \citenamefont {Wu}, \citenamefont {Ye}, \citenamefont {Zhang}, \citenamefont {Wang}, \citenamefont {Lian}, \citenamefont {Yao}, \citenamefont {Xu}, \citenamefont {Zhang}, \citenamefont {Xu} \emph {et~al.}}]{HL2}%
  \BibitemOpen
  \bibfield  {author} {\bibinfo {author} {\bibfnamefont {S.-A.}\ \bibnamefont {Guo}}, \bibinfo {author} {\bibfnamefont {Y.-K.}\ \bibnamefont {Wu}}, \bibinfo {author} {\bibfnamefont {J.}~\bibnamefont {Ye}}, \bibinfo {author} {\bibfnamefont {L.}~\bibnamefont {Zhang}}, \bibinfo {author} {\bibfnamefont {Y.}~\bibnamefont {Wang}}, \bibinfo {author} {\bibfnamefont {W.-Q.}\ \bibnamefont {Lian}}, \bibinfo {author} {\bibfnamefont {R.}~\bibnamefont {Yao}}, \bibinfo {author} {\bibfnamefont {Y.-L.}\ \bibnamefont {Xu}}, \bibinfo {author} {\bibfnamefont {C.}~\bibnamefont {Zhang}}, \bibinfo {author} {\bibfnamefont {Y.-Z.}\ \bibnamefont {Xu}},  \emph {et~al.},\ }\href@noop {} {\bibfield  {journal} {\bibinfo  {journal} {Science Advances}\ }\textbf {\bibinfo {volume} {11}},\ \bibinfo {pages} {eadt4713} (\bibinfo {year} {2025})}\BibitemShut {NoStop}%
\bibitem [{\citenamefont {Dutt}\ \emph {et~al.}(2021)\citenamefont {Dutt}, \citenamefont {Pednault}, \citenamefont {Wu}, \citenamefont {Sheldon}, \citenamefont {Smolin}, \citenamefont {Bishop},\ and\ \citenamefont {Chuang}}]{HL3}%
  \BibitemOpen
  \bibfield  {author} {\bibinfo {author} {\bibfnamefont {A.}~\bibnamefont {Dutt}}, \bibinfo {author} {\bibfnamefont {E.}~\bibnamefont {Pednault}}, \bibinfo {author} {\bibfnamefont {C.~W.}\ \bibnamefont {Wu}}, \bibinfo {author} {\bibfnamefont {S.}~\bibnamefont {Sheldon}}, \bibinfo {author} {\bibfnamefont {J.}~\bibnamefont {Smolin}}, \bibinfo {author} {\bibfnamefont {L.}~\bibnamefont {Bishop}}, \ and\ \bibinfo {author} {\bibfnamefont {I.~L.}\ \bibnamefont {Chuang}},\ }\href@noop {} {\bibfield  {journal} {\bibinfo  {journal} {URL https://arxiv. org/abs/2112.14553}\ } (\bibinfo {year} {2021})}\BibitemShut {NoStop}%
\bibitem [{\citenamefont {Huang}\ \emph {et~al.}(2023)\citenamefont {Huang}, \citenamefont {Tong}, \citenamefont {Fang},\ and\ \citenamefont {Su}}]{Hsin2023}%
  \BibitemOpen
  \bibfield  {author} {\bibinfo {author} {\bibfnamefont {H.-Y.}\ \bibnamefont {Huang}}, \bibinfo {author} {\bibfnamefont {Y.}~\bibnamefont {Tong}}, \bibinfo {author} {\bibfnamefont {D.}~\bibnamefont {Fang}}, \ and\ \bibinfo {author} {\bibfnamefont {Y.}~\bibnamefont {Su}},\ }\href@noop {} {\bibfield  {journal} {\bibinfo  {journal} {Physical Review Letters}\ }\textbf {\bibinfo {volume} {130}},\ \bibinfo {pages} {200403} (\bibinfo {year} {2023})}\BibitemShut {NoStop}%
\bibitem [{\citenamefont {Bakshi}\ \emph {et~al.}()\citenamefont {Bakshi}, \citenamefont {Liu}, \citenamefont {Moitra},\ and\ \citenamefont {Tang}}]{Ainesh2024}%
  \BibitemOpen
  \bibfield  {author} {\bibinfo {author} {\bibfnamefont {A.}~\bibnamefont {Bakshi}}, \bibinfo {author} {\bibfnamefont {A.}~\bibnamefont {Liu}}, \bibinfo {author} {\bibfnamefont {A.}~\bibnamefont {Moitra}}, \ and\ \bibinfo {author} {\bibfnamefont {E.}~\bibnamefont {Tang}},\ }\href@noop {} {\bibinfo  {journal} {URL https://arxiv. org/abs/2405.00082}\ }\BibitemShut {NoStop}%
\bibitem [{\citenamefont {Hu}\ \emph {et~al.}(2025)\citenamefont {Hu}, \citenamefont {Ma}, \citenamefont {Gong}, \citenamefont {Ye}, \citenamefont {Tong}, \citenamefont {Flammia},\ and\ \citenamefont {Yelin}}]{Hu2025}%
  \BibitemOpen
\bibfield  {journal} {  }\bibfield  {author} {\bibinfo {author} {\bibfnamefont {H.-Y.}\ \bibnamefont {Hu}}, \bibinfo {author} {\bibfnamefont {M.}~\bibnamefont {Ma}}, \bibinfo {author} {\bibfnamefont {W.}~\bibnamefont {Gong}}, \bibinfo {author} {\bibfnamefont {Q.}~\bibnamefont {Ye}}, \bibinfo {author} {\bibfnamefont {Y.}~\bibnamefont {Tong}}, \bibinfo {author} {\bibfnamefont {S.~T.}\ \bibnamefont {Flammia}}, \ and\ \bibinfo {author} {\bibfnamefont {S.~F.}\ \bibnamefont {Yelin}},\ }\href@noop {} {\bibfield  {journal} {\bibinfo  {journal} {arXiv preprint arXiv:2502.11900}\ } (\bibinfo {year} {2025})}\BibitemShut {NoStop}%
\bibitem [{\citenamefont {Li}\ \emph {et~al.}(2024)\citenamefont {Li}, \citenamefont {Tong}, \citenamefont {Gefen}, \citenamefont {Ni},\ and\ \citenamefont {Ying}}]{Haoya2023}%
  \BibitemOpen
  \bibfield  {author} {\bibinfo {author} {\bibfnamefont {H.}~\bibnamefont {Li}}, \bibinfo {author} {\bibfnamefont {Y.}~\bibnamefont {Tong}}, \bibinfo {author} {\bibfnamefont {T.}~\bibnamefont {Gefen}}, \bibinfo {author} {\bibfnamefont {H.}~\bibnamefont {Ni}}, \ and\ \bibinfo {author} {\bibfnamefont {L.}~\bibnamefont {Ying}},\ }\href@noop {} {\bibfield  {journal} {\bibinfo  {journal} {npj Quantum Information}\ }\textbf {\bibinfo {volume} {10}},\ \bibinfo {pages} {83} (\bibinfo {year} {2024})}\BibitemShut {NoStop}%
\bibitem [{\citenamefont {Mirani}\ and\ \citenamefont {Hayden}()}]{Arjun2024}%
  \BibitemOpen
  \bibfield  {author} {\bibinfo {author} {\bibfnamefont {A.}~\bibnamefont {Mirani}}\ and\ \bibinfo {author} {\bibfnamefont {P.}~\bibnamefont {Hayden}},\ }\href@noop {} {\bibinfo  {journal} {arXiv preprint arXiv:2403.00069}\ }\BibitemShut {NoStop}%
\bibitem [{\citenamefont {Braak}(2011)}]{Rabi}%
  \BibitemOpen
\bibfield  {journal} {  }\bibfield  {author} {\bibinfo {author} {\bibfnamefont {D.}~\bibnamefont {Braak}},\ }\href@noop {} {\bibfield  {journal} {\bibinfo  {journal} {Physical Review Letters}\ }\textbf {\bibinfo {volume} {107}},\ \bibinfo {pages} {100401} (\bibinfo {year} {2011})}\BibitemShut {NoStop}%
\bibitem [{sup()}]{supmat}%
  \BibitemOpen
  \href@noop {} {\enquote {\bibinfo {title} {Supplementary material},}\ }\bibinfo {howpublished} {\url{URL_will_be_inserted_by_publisher}}\BibitemShut {NoStop}%
\bibitem [{\citenamefont {Kimmel}\ \emph {et~al.}(2015)\citenamefont {Kimmel}, \citenamefont {Low},\ and\ \citenamefont {Yoder}}]{Shelby2015}%
  \BibitemOpen
  \bibfield  {author} {\bibinfo {author} {\bibfnamefont {S.}~\bibnamefont {Kimmel}}, \bibinfo {author} {\bibfnamefont {G.~H.}\ \bibnamefont {Low}}, \ and\ \bibinfo {author} {\bibfnamefont {T.~J.}\ \bibnamefont {Yoder}},\ }\href@noop {} {\bibfield  {journal} {\bibinfo  {journal} {Physical Review A}\ }\textbf {\bibinfo {volume} {92}},\ \bibinfo {pages} {062315} (\bibinfo {year} {2015})}\BibitemShut {NoStop}%
\bibitem [{\citenamefont {Zhuang}\ \emph {et~al.}(2018)\citenamefont {Zhuang}, \citenamefont {Zhang},\ and\ \citenamefont {Shapiro}}]{Quntao2018}%
  \BibitemOpen
  \bibfield  {author} {\bibinfo {author} {\bibfnamefont {Q.}~\bibnamefont {Zhuang}}, \bibinfo {author} {\bibfnamefont {Z.}~\bibnamefont {Zhang}}, \ and\ \bibinfo {author} {\bibfnamefont {J.~H.}\ \bibnamefont {Shapiro}},\ }\href@noop {} {\bibfield  {journal} {\bibinfo  {journal} {Physical Review A}\ }\textbf {\bibinfo {volume} {97}},\ \bibinfo {pages} {032329} (\bibinfo {year} {2018})}\BibitemShut {NoStop}%
\bibitem [{\citenamefont {Zhao}\ \emph {et~al.}(2023)\citenamefont {Zhao}, \citenamefont {Han}, \citenamefont {Kivelson},\ and\ \citenamefont {Esterlis}}]{Holstein_model}%
  \BibitemOpen
  \bibfield  {author} {\bibinfo {author} {\bibfnamefont {S.}~\bibnamefont {Zhao}}, \bibinfo {author} {\bibfnamefont {Z.}~\bibnamefont {Han}}, \bibinfo {author} {\bibfnamefont {S.~A.}\ \bibnamefont {Kivelson}}, \ and\ \bibinfo {author} {\bibfnamefont {I.}~\bibnamefont {Esterlis}},\ }\href@noop {} {\bibfield  {journal} {\bibinfo  {journal} {Physical Review B}\ }\textbf {\bibinfo {volume} {107}},\ \bibinfo {pages} {075142} (\bibinfo {year} {2023})}\BibitemShut {NoStop}%
\bibitem [{\citenamefont {Shih}\ and\ \citenamefont {Berkelbach}(2024)}]{SSH_model}%
  \BibitemOpen
  \bibfield  {author} {\bibinfo {author} {\bibfnamefont {P.}~\bibnamefont {Shih}}\ and\ \bibinfo {author} {\bibfnamefont {T.~C.}\ \bibnamefont {Berkelbach}},\ }\href@noop {} {\bibfield  {journal} {\bibinfo  {journal} {The Journal of Chemical Physics}\ }\textbf {\bibinfo {volume} {160}} (\bibinfo {year} {2024})}\BibitemShut {NoStop}%
\bibitem [{\citenamefont {Bursill}\ \emph {et~al.}(1999)\citenamefont {Bursill}, \citenamefont {McKenzie},\ and\ \citenamefont {Hamer}}]{SPmodel}%
  \BibitemOpen
  \bibfield  {author} {\bibinfo {author} {\bibfnamefont {R.~J.}\ \bibnamefont {Bursill}}, \bibinfo {author} {\bibfnamefont {R.~H.}\ \bibnamefont {McKenzie}}, \ and\ \bibinfo {author} {\bibfnamefont {C.~J.}\ \bibnamefont {Hamer}},\ }\href@noop {} {\bibfield  {journal} {\bibinfo  {journal} {Physical review letters}\ }\textbf {\bibinfo {volume} {83}},\ \bibinfo {pages} {408} (\bibinfo {year} {1999})}\BibitemShut {NoStop}%
\bibitem [{\citenamefont {Buyens}\ \emph {et~al.}(2016)\citenamefont {Buyens}, \citenamefont {Verstraete},\ and\ \citenamefont {Van~Acoleyen}}]{Schwinger_model}%
  \BibitemOpen
  \bibfield  {author} {\bibinfo {author} {\bibfnamefont {B.}~\bibnamefont {Buyens}}, \bibinfo {author} {\bibfnamefont {F.}~\bibnamefont {Verstraete}}, \ and\ \bibinfo {author} {\bibfnamefont {K.}~\bibnamefont {Van~Acoleyen}},\ }\href@noop {} {\bibfield  {journal} {\bibinfo  {journal} {Physical Review D}\ }\textbf {\bibinfo {volume} {94}},\ \bibinfo {pages} {085018} (\bibinfo {year} {2016})}\BibitemShut {NoStop}%
\bibitem [{\citenamefont {Liu}\ \emph {et~al.}(2024)\citenamefont {Liu}, \citenamefont {Singh}, \citenamefont {Smith}, \citenamefont {Crane}, \citenamefont {Martyn}, \citenamefont {Eickbusch}, \citenamefont {Schuckert}, \citenamefont {Li}, \citenamefont {Sinanan-Singh}, \citenamefont {Soley} \emph {et~al.}}]{liu2024hybrid}%
  \BibitemOpen
  \bibfield  {author} {\bibinfo {author} {\bibfnamefont {Y.}~\bibnamefont {Liu}}, \bibinfo {author} {\bibfnamefont {S.}~\bibnamefont {Singh}}, \bibinfo {author} {\bibfnamefont {K.~C.}\ \bibnamefont {Smith}}, \bibinfo {author} {\bibfnamefont {E.}~\bibnamefont {Crane}}, \bibinfo {author} {\bibfnamefont {J.~M.}\ \bibnamefont {Martyn}}, \bibinfo {author} {\bibfnamefont {A.}~\bibnamefont {Eickbusch}}, \bibinfo {author} {\bibfnamefont {A.}~\bibnamefont {Schuckert}}, \bibinfo {author} {\bibfnamefont {R.~D.}\ \bibnamefont {Li}}, \bibinfo {author} {\bibfnamefont {J.}~\bibnamefont {Sinanan-Singh}}, \bibinfo {author} {\bibfnamefont {M.~B.}\ \bibnamefont {Soley}},  \emph {et~al.},\ }\href@noop {} {\bibfield  {journal} {\bibinfo  {journal} {arXiv preprint arXiv:2407.10381}\ } (\bibinfo {year} {2024})}\BibitemShut {NoStop}%
\bibitem [{\citenamefont {Safavi-Naini}\ \emph {et~al.}(2018)\citenamefont {Safavi-Naini}, \citenamefont {Lewis-Swan}, \citenamefont {Bohnet}, \citenamefont {G{\"a}rttner}, \citenamefont {Gilmore}, \citenamefont {Jordan}, \citenamefont {Cohn}, \citenamefont {Freericks}, \citenamefont {Rey},\ and\ \citenamefont {Bollinger}}]{safavi2018verification}%
  \BibitemOpen
  \bibfield  {author} {\bibinfo {author} {\bibfnamefont {A.}~\bibnamefont {Safavi-Naini}}, \bibinfo {author} {\bibfnamefont {R.}~\bibnamefont {Lewis-Swan}}, \bibinfo {author} {\bibfnamefont {J.~G.}\ \bibnamefont {Bohnet}}, \bibinfo {author} {\bibfnamefont {M.}~\bibnamefont {G{\"a}rttner}}, \bibinfo {author} {\bibfnamefont {K.}~\bibnamefont {Gilmore}}, \bibinfo {author} {\bibfnamefont {J.}~\bibnamefont {Jordan}}, \bibinfo {author} {\bibfnamefont {J.}~\bibnamefont {Cohn}}, \bibinfo {author} {\bibfnamefont {J.~K.}\ \bibnamefont {Freericks}}, \bibinfo {author} {\bibfnamefont {A.~M.}\ \bibnamefont {Rey}}, \ and\ \bibinfo {author} {\bibfnamefont {J.}~\bibnamefont {Bollinger}},\ }\href@noop {} {\bibfield  {journal} {\bibinfo  {journal} {Physical review letters}\ }\textbf {\bibinfo {volume} {121}},\ \bibinfo {pages} {040503} (\bibinfo {year} {2018})}\BibitemShut {NoStop}%
\bibitem [{\citenamefont {Magazz{\`u}}\ \emph {et~al.}(2018)\citenamefont {Magazz{\`u}}, \citenamefont {Forn-D{\'\i}az}, \citenamefont {Belyansky}, \citenamefont {Orgiazzi}, \citenamefont {Yurtalan}, \citenamefont {Otto}, \citenamefont {Lupascu}, \citenamefont {Wilson},\ and\ \citenamefont {Grifoni}}]{magazzu2018probing}%
  \BibitemOpen
  \bibfield  {author} {\bibinfo {author} {\bibfnamefont {L.}~\bibnamefont {Magazz{\`u}}}, \bibinfo {author} {\bibfnamefont {P.}~\bibnamefont {Forn-D{\'\i}az}}, \bibinfo {author} {\bibfnamefont {R.}~\bibnamefont {Belyansky}}, \bibinfo {author} {\bibfnamefont {J.-L.}\ \bibnamefont {Orgiazzi}}, \bibinfo {author} {\bibfnamefont {M.}~\bibnamefont {Yurtalan}}, \bibinfo {author} {\bibfnamefont {M.~R.}\ \bibnamefont {Otto}}, \bibinfo {author} {\bibfnamefont {A.}~\bibnamefont {Lupascu}}, \bibinfo {author} {\bibfnamefont {C.}~\bibnamefont {Wilson}}, \ and\ \bibinfo {author} {\bibfnamefont {M.}~\bibnamefont {Grifoni}},\ }\href@noop {} {\bibfield  {journal} {\bibinfo  {journal} {Nature communications}\ }\textbf {\bibinfo {volume} {9}},\ \bibinfo {pages} {1403} (\bibinfo {year} {2018})}\BibitemShut {NoStop}%
\bibitem [{\citenamefont {De~Filippis}\ \emph {et~al.}(2020)\citenamefont {De~Filippis}, \citenamefont {De~Candia}, \citenamefont {Cangemi}, \citenamefont {Sassetti}, \citenamefont {Fazio},\ and\ \citenamefont {Cataudella}}]{de2020quantum}%
  \BibitemOpen
  \bibfield  {author} {\bibinfo {author} {\bibfnamefont {G.}~\bibnamefont {De~Filippis}}, \bibinfo {author} {\bibfnamefont {A.}~\bibnamefont {De~Candia}}, \bibinfo {author} {\bibfnamefont {L.}~\bibnamefont {Cangemi}}, \bibinfo {author} {\bibfnamefont {M.}~\bibnamefont {Sassetti}}, \bibinfo {author} {\bibfnamefont {R.}~\bibnamefont {Fazio}}, \ and\ \bibinfo {author} {\bibfnamefont {V.}~\bibnamefont {Cataudella}},\ }\href@noop {} {\bibfield  {journal} {\bibinfo  {journal} {Physical Review B}\ }\textbf {\bibinfo {volume} {101}},\ \bibinfo {pages} {180408} (\bibinfo {year} {2020})}\BibitemShut {NoStop}%
\bibitem [{\citenamefont {Zueco}\ \emph {et~al.}(2008)\citenamefont {Zueco}, \citenamefont {H{\"a}nggi},\ and\ \citenamefont {Kohler}}]{Spec_den}%
  \BibitemOpen
  \bibfield  {author} {\bibinfo {author} {\bibfnamefont {D.}~\bibnamefont {Zueco}}, \bibinfo {author} {\bibfnamefont {P.}~\bibnamefont {H{\"a}nggi}}, \ and\ \bibinfo {author} {\bibfnamefont {S.}~\bibnamefont {Kohler}},\ }\href@noop {} {\bibfield  {journal} {\bibinfo  {journal} {New Journal of Physics}\ }\textbf {\bibinfo {volume} {10}},\ \bibinfo {pages} {115012} (\bibinfo {year} {2008})}\BibitemShut {NoStop}%
\bibitem [{\citenamefont {Peng}\ \emph {et~al.}(2024)\citenamefont {Peng}, \citenamefont {Liu}, \citenamefont {Chern},\ and\ \citenamefont {Luo}}]{peng2024provably}%
  \BibitemOpen
  \bibfield  {author} {\bibinfo {author} {\bibfnamefont {C.}~\bibnamefont {Peng}}, \bibinfo {author} {\bibfnamefont {J.-P.}\ \bibnamefont {Liu}}, \bibinfo {author} {\bibfnamefont {G.-W.}\ \bibnamefont {Chern}}, \ and\ \bibinfo {author} {\bibfnamefont {D.}~\bibnamefont {Luo}},\ }\href@noop {} {\bibfield  {journal} {\bibinfo  {journal} {arXiv preprint arXiv:2408.00276}\ } (\bibinfo {year} {2024})}\BibitemShut {NoStop}%
\end{thebibliography}%

\bibliographystyle{apsrev4-1}

\appendix

\clearpage

\onecolumngrid
\begin{center}
	\noindent\textbf{Supplementary Material}
	\bigskip
		
	\noindent\textbf{\large{}}
\end{center}

\onecolumngrid

\section{Random Unitary Tranformations}
Here, we provide detailed description of the random unitary transformation (RUT) algorithm. Let ${\hat H}$ be arbitrary Hamiltonian, and $\tau \equiv t/R$, we have:
\begin{eqnarray}\label{sup_insert}
e^{-i{\hat H}^{(R)}t} = \prod_{i}^{R}{\mathbb U}^\dagger(\boldsymbol \theta_i) e^{-i{\hat H}\tau} {\mathbb U}(\boldsymbol \theta_i) \approx   {\rm exp}[-it \sum_{i=1}^{R}{\mathbb U}^\dagger(\boldsymbol \theta_i){\hat H}{\mathbb U}(\boldsymbol \theta_i)] \approx  {\rm exp}[-\frac{it}{ V(\boldsymbol{\theta})}\int_{\Omega} d{\boldsymbol{\theta}} {\mathbb U}^\dagger(\boldsymbol \theta){\hat H}{\mathbb U}(\boldsymbol \theta)
\end{eqnarray}
where ${\boldsymbol{\theta}} \equiv (\theta_1, \theta_2, ..., \theta_j)$ represents all $\theta_j$ that parameterize the unitary sequence ${\mathbb U}$. $\theta_j$ are sampled from independent uniform distribution $\mathcal{U}_j$, and $\Omega$ is the sampling domain of $\boldsymbol{\theta}$ with volume $V(\boldsymbol{\theta})$. We define the effective Hamiltonian at the limit $R \rightarrow \infty$:
\begin{eqnarray}
\hat{\mathcal H} \equiv {\hat H}^{(R=\infty)} = 
\frac{1}{V(\boldsymbol{\theta}) }\int d{\boldsymbol{\theta}} {\mathbb U}^\dagger(\boldsymbol \theta){\hat H}{\mathbb U}(\boldsymbol \theta)
 \end{eqnarray}
However, when $R \neq \infty$, deviation occurs between ${\hat H}^{(R)}$ and $\hat{\mathcal H}$. The deviation can be contributed to two types of error. At the first approximation sign in Eq.~\ref{sup_insert}, trotter error is introduced as the commutator $[{\mathbb U}(\boldsymbol{\theta}_{i1}),{\mathbb U}(\boldsymbol{\theta}_{i2})] \neq 0$. At the second approximation sign in Eq.~\ref{sup_insert}, Monte-Carlo error is introduced by approximating the continuous integral with a finite number of samples. However, when $R \rightarrow \infty$, the both of the errors vanish.

By choosing different forms of ${\mathbb U}(\boldsymbol \theta)$, different terms in ${\hat H}$ can be integrated out in $\hat{\mathcal H}$ based on their symmetry. In the manuscript, the first unitary sequence used in the reshaping process is $\mathbb{U}^{(1)} = \prod_n^{N_b} e^{-i \theta {\hat b}_n^\dagger {\hat b}_n}$. The reshaping mechanism of $\mathbb{U}^{(1)}$ relies on the following equation~\cite{Haoya2023}:
\begin{eqnarray}\label{boson random}
e^{i \theta {\hat b}_n^\dagger {\hat b}_n} {\hat b_n} e^{-i \theta {\hat b}_n^\dagger {\hat b}_n} = e^{-i\theta}{\hat b_n},~~~ e^{i \theta {\hat b}_n^\dagger {\hat b}_n} {\hat b_n^\dagger} e^{-i \theta {\hat b}_n^\dagger {\hat b}_n} = e^{i\theta}{\hat b_n}
\end{eqnarray}
which be derived by applying $e^{i \theta {\hat b}_n^\dagger {\hat b}_n} {\hat b_n} e^{-i \theta {\hat b}_n^\dagger {\hat b}_n}$ on arbitrary fock state $\ket{n}$. For example, for ${\hat b}_n$ we have:
\begin{eqnarray}
e^{i \theta {\hat b}_n^\dagger {\hat b}_n} {\hat b_n} e^{-i \theta {\hat b}_n^\dagger {\hat b}_n}\ket{m} = e^{i \theta (m-1)}\sqrt{m}e^{-i \theta m}\ket{m-1} = e^{-i\theta}\sqrt{m}\ket{m-1}  = e^{-i\theta}{\hat b}_n\ket{m}
\end{eqnarray}
If we choose $\theta$ from uniform distribution $\mathcal{U}(0,2\pi)$, terms in ${\hat H}$ that contains ${\hat b}_n$ or ${\hat b}_n^\dagger$ will be canceled out, as $\int_0^{2\pi} e^{\pm i \theta} d\theta = 0$.

The second unitary used in the manuscript is:
\begin{eqnarray}\label{pauli_random_supp}
\mathbb{U}^{(2)}_b = \prod_{j}^{N_q} U_j~~{\rm with}~~ U_{j}= 
    \begin{cases}
        e^{-i\theta_{j}\mathcal{P}_j^{b}} & \text{if } j \in {\rm supp}({\hat E}_{b})\\
        e^{-i\theta_{j}\mathcal{P}_j }e^{-i\phi_j\mathcal{P}_j^\prime} & \text{if } j \not\in {\rm supp}({\hat E}_{b})
    \end{cases}
\end{eqnarray}
where ${\hat E}_b$ is a pre-selected Pauli string with $|{\rm supp}({\hat E}_{b})| = k$. $\mathcal{P}_j^{b}$ refers to the $j^{th}$ operator of Pauli string $E_{b}$. $\mathcal{P}_j$ and $\mathcal{P}_j^{\prime}$ are arbitrary Pauli operators satisfying $[\mathcal{P}_j, \mathcal{P}_j^{\prime}] \neq 0$.  The reshaping mechanism of $\mathbb{U}^{(2)}_b$ relies on the commutation relationship of Pauli operators. Consider the action of $e^{-i\theta_{j}\mathcal{P}_j^{b}}$ on an arbitrary Pauli operator $\mathcal{P}_j$, we have:
\begin{eqnarray}\label{pauli random}
e^{i\theta_{j}\mathcal{P}_j^{b}} \mathcal{P}_j e^{-i\theta_{j}\mathcal{P}_j^{b}} = 
    \begin{cases}
        [e^{2i\theta_j}(\mathcal{P}_j + \epsilon \mathcal{P}_j^{\prime}) + e^{-2i\theta_j}(\mathcal{P}_j - \epsilon \mathcal{P}_j^{\prime})]/2 & \text{if } \mathcal{P}_j^{b} \neq  \mathcal{P}_j\\
        \mathcal{P}_j^{b} & \text{if } \mathcal{P}_j^{b} =  \mathcal{P}_j
    \end{cases}
\end{eqnarray}
where $\epsilon \equiv \epsilon({\mathcal{P}_j^b, \mathcal{P}_j, \mathcal{P}_j^{\prime}})$ is the Levi-Civita symbol, where we choose $\{\mathcal{P}_j^b, = 1; \mathcal{P}_j = 2; \mathcal{P}_j^{\prime} = 3\}$. Again, let $\theta_j \sim \mathcal{U}_j(0, \pi)$, any terms in $\hat H$ that contains $\mathcal{P}_j$ will be canceled out unless $\mathcal{P}_j^b =  \mathcal{P}_j$. As a concrete example, consider the following 3-qubit Hamiltonian:
\begin{eqnarray}
{\hat H}_s = ZII + ZIX + YIX + ZZX
\end{eqnarray}
we choose ${\hat E}_b = ZIX$, and construct $\mathbb{U}^{(2)}_b = e^{-i\theta_1Z_1}e^{-i(\theta_2X_2 + \phi_2Y_2) }e^{-i\theta_3X_3}$ according to Eq.~\ref{pauli_random_supp}. Let $\boldsymbol{\theta} = (\theta_1, \theta_2, \phi_2, \theta_3)$ and $\boldsymbol{\theta} \sim \mathcal{U}(0,\pi)^{4}$, reshaping ${\hat H}_s$ with $\mathbb{U}^{(2)}_b$ gives:
\begin{eqnarray}
{\mathcal{\hat H}}_s &=& \frac{1}{V(\boldsymbol{\theta}) }\int d{\boldsymbol{\theta}} [\mathbb{U}^{(2)}_b(\boldsymbol \theta)]^\dagger{\hat H}_s\mathbb{U}^{(2)}_b (\boldsymbol \theta) \nonumber \\
&=& \frac{1}{V(\boldsymbol{\theta}) }\int d{\boldsymbol{\theta}} \Big\{ZII +ZIX + [{\rm cos}(2\theta_1)YIX + {\rm sin}(2\theta_1)XIX]\Big\}\nonumber \\
&&+ \frac{1}{V(\boldsymbol{\theta}) }\int d{\boldsymbol{\theta}} \Big\{e^{i\theta_2X_2}[{\rm cos}(2\phi_2)ZZX - {\rm sin}(2\phi_2)ZXX]e^{-i\theta_2X_2}\Big\} \nonumber \\
&=& \frac{1}{V(\boldsymbol{\theta}) }\int d{\boldsymbol{\theta}} \Big\{ZII +ZIX + [{\rm cos}(2\theta_1)YIX + {\rm sin}(2\theta_1)XIX]\Big\}\nonumber \\
&&+\frac{1}{V(\boldsymbol{\theta}) }\int d{\boldsymbol{\theta}} \Big\{{\rm cos}(2\phi_2)[{\rm cos}(2\theta_2)ZZX+{\rm sin}(2\theta_2)ZYX] - {\rm sin}(2\phi_2)ZXX\Big\} \nonumber \\
&=& ZII + ZIX + 0 + 0
\end{eqnarray}

Therefore, the collective behavior of $U_j$ acting on an arbitrary Pauli string ${\hat E}_a$ can be concluded as:
\begin{enumerate}
  \item If $j \in {\rm supp}({\hat E}_{b})$ and  $\mathcal{P}_j^a = \mathcal{P}_j^{b}/ \mathcal{I}$, ${\hat E}_a$ is kept as $U_j^\dagger \mathcal{P}_j^a U_j = \mathcal{P}_j^a$. 
  \item If $j \in {\rm supp}({\hat E}_{b})$ and  $\mathcal{P}_j^a \neq \mathcal{P}_j^{b}/ \mathcal{I}$, ${\hat E}_a$ is canceled due to the extra phase factor.
  \item If $j \not\in {\rm supp}({\hat E}_{b})$, ${\hat E}_a$ is canceled unless $\mathcal{P}_j^a = \mathcal{I}$
\end{enumerate}
As $\mathbb{U}^{(2)}_b$ includes $U_{j}$ across all qubits, reshaping the spin-part of $\hat{\mathcal H}^{(1)}$ with $\mathbb{U}^{(2)}_b$ leads to $\hat{\mathcal H}_{\rm S}^{(2)}$, where only Pauli strings in set $S_b$ are kept in the effective Hamiltonian.

\section{Estimation error propagation}
As described in the manuscript, we use ${\tilde \omega}_n$ as the input of the trotter-based scheme and DQS-based scheme. However, ${\tilde \omega}_n$ deviates from its actual value $\omega_n$ by $\epsilon({ \omega}_n)$, which leads to the propagation of error in our learning protocol. Here, we prove that this error can be suppressed while maintaining the Heisenberg limit. The effective time evolution of the bosonic wavefunction is: 
\begin{eqnarray}\label{sup_displacement}
\ket{\Psi_{\rm B}(t)}_l =  \prod_n{\hat D}_{n}(\frac{\Lambda_{b,l}^n}{\omega_{n}}(e^{-i\omega_{n} t} - 1)) \ket{\Psi_{\rm B}(0)} 
\end{eqnarray}

For the trotter-based scheme, we insert $\mathbb{U}^{(3)}$ during the reshaping process to cancel $\omega_n$. After canceling $\omega_n$ with ${\tilde \omega}_n$, the error $\epsilon({ \omega}_n)$ can be analyzed by swapping $\omega_n$ with $\epsilon({ \omega}_n) = |{\omega}_n - {\tilde \omega}_n|$ in Eq.~\ref{sup_displacement}. Taylor expansion of the error term leads to:
\begin{eqnarray}
\ket{\Psi_{\rm B}^{\rm eff}(t)}_l &=&  \prod_n{\hat D}_{n}(\frac{\Lambda_{b,l}^n}{\epsilon({ \omega}_n)}(e^{-i\epsilon({ \omega}_n) t} - 1)) \ket{\bf 0} \nonumber \\
&=&\prod_n{\hat D}_{n}(\frac{\Lambda_{b,l}^n}{\epsilon({ \omega}_n)}(-i\epsilon({ \omega}_n) t + \mathcal{R}_2[\epsilon(\omega_n)]) \ket{\bf 0} \nonumber \\
&=& \prod_n{\hat D}_{n}(-i\Lambda_{b,l}^n t + \frac{\Lambda_{b,l}^n\mathcal{R}_2[\epsilon({ \omega}_n)]}{\epsilon({ \omega}_n)}) \ket{\bf 0}
\end{eqnarray}
where $\mathcal{R}_2[\epsilon({ \omega}_n)]$ is the remainder for the first-order Taylor expansion. As $\mathcal{R}_2[\epsilon({ \omega}_n)] \sim \mathcal{O}[\epsilon({ \omega}_n)^2]$, we have $\frac{\Lambda_{b,l}^n\mathcal{R}_2[\epsilon({ \omega}_n)]}{\epsilon({ \omega}_n)} \sim \mathcal{O}[\epsilon({ \omega}_n)^1]$. The homodyne measurement on the momentum quadrature is $\langle \hat P_n \rangle = -\sqrt{2}\Lambda_{b,l}^n t + \mathcal{O}[\epsilon({ \omega}_n)]$, which includes the error introduced by ${\tilde \omega}_n$. However, RFE tolerates a maximum failure probability $\delta_{\rm max}$ inherently\cite{Haoya2023}. Suppose the maximum evolution time used in RFE is $2^\mathfrak{K^\ast}$, as long as the deviation originated from $\epsilon({ \omega}_n)$ is smaller than $\delta_{\rm max}(\mathfrak{K^\ast})$ at $t = 2^\mathfrak{K^\ast}$ (which can be achieved by decreasing $\epsilon({ \omega}_n)$), RFE can be successfully implemented. Although higher $\epsilon({ \omega}_n)$ would decrease the SPAM error tolerance of the trotter-based scheme, it still achieves the Heisenberg limit.

For the DQS-based scheme, we measure the entangled wavefunction at $t = \pi/{\tilde \omega}_n$. Start from Eq.~\ref{Supp_Entangled_displacement}, we swap $t$ with $\pi/{\tilde \omega}_n$. Taylor expansion on error term leads to:
\begin{eqnarray}
\ket{\Psi_{\rm B}^{\rm ent}(\pi/{\tilde \omega}_n)}_l &=&  \prod_n{\hat D}_{n}(\frac{\Lambda_{b,l}^n}{\omega_{n}}(e^{-i\pi \omega_n/{\tilde \omega}_n} - 1)) S_{n}^{(W)}(e^{-2i\pi \omega_n/{\tilde \omega}_n}z) \ket{\bf 0}  \nonumber \\
&=&  \prod_n{\hat D}_{n}(\frac{\Lambda_{b,l}^n}{\omega_{n}}(-1-e^{-i\pi \epsilon({ \omega}_n)/{\tilde \omega}_n} )) S_{n}^{(W)}(e^{-2i\pi \epsilon({ \omega}_n)/{\tilde \omega}_n}z)\ket{\bf 0} \nonumber \\
&=&  \prod_n{\hat D}_{n}(\frac{\Lambda_{b,l}^n}{\omega_{n}}(-2+\mathcal{O}[\epsilon({ \omega}_n)])) S_{n}^{(W)}(z+\mathcal{O}[\epsilon({ \omega}_n)])\ket{\bf 0} \nonumber \\
\end{eqnarray}
where ${\hat S}_{n}^{(W)} \equiv {\rm exp}[\frac{1}{2}(z^\ast {\hat B}_{n}^2 - z {\hat B}_{n}^{\dagger 2})]$ is the entangled squeezing operator with ${\hat B}_n \equiv \sum_{w=1}^{W} {\hat b_{n, w}} / \sqrt{W}$.  As error in squeezing parameter does not affect the mean value of displacement for the squeezed state, choosing ${\tilde{\mathcal X}}_n \equiv \sum_{w=1}^W \langle{\hat X_{n, w}}\rangle/W$ as the displacement estimator leads to
\begin{eqnarray}
\mathbb{E}[{\tilde{\mathcal X}}_n] &=& -\frac{2\sqrt{2}\Lambda_{b,l}^n}{\omega_n} + \mathcal{O}[\epsilon({ \omega}_n)]
\end{eqnarray}
where $\mathcal{O}[\epsilon({ \omega}_n)]$ is a biased error. Inverting the above equation gives:$\Lambda_{b,l}^n = -\frac{\omega_n{\tilde{\mathcal X}}_n}{2\sqrt{2}} + \mathcal{O}[\epsilon({ \omega}_n)]$. As $\epsilon({ \omega}_n) \sim \mathcal{O}(T^{-1})$, we can conclude that $\epsilon(\Lambda_{b,l}^n) \sim \mathcal{O}(T^{-1})$, which achieves the Heisenberg limit.

\section{Extraction of $\Xi_{b,l}$ using robust phase estimation}
The general Hamiltonian considered in this paper is:
\begin{eqnarray}\label{sup_genH}
{\hat H} = {\hat H}_{\rm spin} + {\hat H}_{\rm boson} + {\hat H}_{\rm int}
\end{eqnarray}
where ${\hat H}_{\rm spin} = \sum_a \xi_a E_a$, ${\hat H}_{b} = \sum_n \omega_n {\hat b}_n^\dagger {\hat b}_n$ and ${\hat H}_{\rm int} = \sum_{n,a} \lambda_a^n E_a({\hat b}_n^\dagger + {\hat b}_n)$.  ${\hat H}_{\rm spin}$ and the spin part of ${\hat H}_{\rm int}$ are considered to be general $k-$local spin Hamiltonians. Therefore, summation over $a$ goes up to $4^k{{N_q}\choose{k}}$. Among them, there are at most $3^k{{N_q}\choose{k}}$ terms with their support equalt to $k$. Therefore, the number of ${\hat E}_{b}$ allowed given $k$ and $N_q$ is $3^k{{N_q}\choose{k}}$. For each ${\hat E}_{b}$, the number of eigenstates is $2^k$, as each Pauli operator has two eigenstates, each with eigenvalue $\pm 1$. We label the eigenstates of ${\hat E}_{b}$ as $\ket{E_{b}}_l$. 

Here, we demonstrate the extraction of $\Xi_{b, l}$ using the robust phase estimation (RPE)~\cite{Shelby2015}. Reshaping ${\hat H}$ with  $\mathbb{U}^{(1)}$ and $\mathbb{U}_b^{(2)}$ gives $\hat{\mathcal H}^{(2)}_{\rm S}$, which takes the form of:
\begin{eqnarray}\label{Eff_Spin_H1}
\hat{\mathcal H}_{\rm S}^{(2)} = \sum_{s: {\hat E}_s \in S_{b}} \xi_s {\hat E}_s ,~~{\rm where}~~ S_{b} = \left\{ \prod_{ i \in \mathrm{supp}({\hat E}_{b})} \mathcal{P}_i^s \;\middle|\; \mathcal{P}_i^s \in \{ \mathcal{P}_i^{b}, \mathcal{I} \} \right\}
\end{eqnarray}
The eigenvalue of $\ket{{\hat E}_{b}}_l$ with respect to $\hat{\mathcal H}^{(2)}_{\rm S}$ is $\Xi_{b,l} \equiv \sum_s \gamma_l^s \xi_s$, where $\gamma_l^s$ is the eigenvalue of $\ket{{\hat E}_{b}}_l$ with respect to ${\hat E}_s$. To implement RPE, we follow~\cite{Hsin2023} and prepare two entangled states that are linear combinations of a pair of $\ket{{\hat E}_{b}}_l$: 
\begin{eqnarray}\label{sup_RPE_state}
\ket{{\Psi}_{\rm ent}^{1}} = \frac{1}{\sqrt{2}}(\ket{{\hat E}_{b}}_{l_1} + \ket{{\hat E}_{b}}_{l_2}) ,~~~\ket{{\Psi}_{\rm ent}^{2}} = \frac{1}{\sqrt{2}}(\ket{{\hat E}_{b}}_{l_1} + i\ket{{\hat E}_{b}}_{l_2})
\end{eqnarray}

Evolving $\ket{{\Psi}_{\rm ent}^{1}}$ and $\ket{{\Psi}_{\rm ent}^{2}}$ in time allows us to measure the phase difference between them. Let $\Delta \Xi \equiv \Xi_{b, l_1} - \Xi_{b, l_2}$, we have:
\begin{eqnarray}
\bra{{\Psi}_{\rm ent}^{1}}  e^{-it \hat{\mathcal H}^{(2)}_{\rm SB}} \ket{{\Psi}_{\rm ent}^{1}} = \frac{1 + {\rm cos}(\Delta \Xi t)}{2}, ~~~\bra{{\Psi}_{\rm ent}^{2}}  e^{-it \hat{\mathcal H}^{(2)}_{\rm SB}} \ket{{\Psi}_{\rm ent}^{2}} = \frac{1 + {\rm sin}(\Delta \Xi t)}{2}
\end{eqnarray}
which allows direct implementation of RPE to find $\Delta \Xi$ at Heisenberg limit. Repeating this procedure for all pairs of $\ket{{\hat E}_{b}}_l$ allows us to solve for all $\Xi_{b, l}$ simultaneously.

\section{Proof of the applicability of robust frequency estimation}
To implement RFE for the trotter-based scheme, we construct a signal $\mathcal{Z}(t) = e^{-i \langle \hat P_n \rangle}$, where $\langle \hat P_n \rangle = -\sqrt{2}{\Lambda}_{b,l}^n t$ with ${\rm Var}[\hat P_n] = 1$.  To prove that RFE can be applied, signal $\mathcal{Z}$ must satisfy the following conditions(~\cite{Haoya2023}, we set the extra phase $f(t) = 0$):
\begin{enumerate}
  \item $|\mathcal{Z}(t)| = 1$ 
  \item $|\mathcal{Z}(t) - e^{-i\langle \hat P_n \rangle}| \le \eta$ with probability at least 1-$\delta$
  \item Generating such $\mathcal{Z}(t)$ requires a evolution time of $\mathcal{O}[t( {\rm log}(\delta^{-1})]$
\end{enumerate}
As the first condition is straightforward, we here prove that the second and the third condition are satisfied. First we define $\tilde{P_n}$, which represents the empirical average of measurement results for $P_n$ over $M$ measurement. Consider a failure probability $\delta$ such that ${\Pr}(|\tilde{ P_n} - \langle \hat P_n \rangle|\ge \eta) = \delta$. Assume $\tilde{P_n}$ follows a normal distribution with $\mu = \langle \hat P_n \rangle$ and $\sigma^2 = 1/M$ (as a result of central limit theorem), we have:
\begin{eqnarray}
\delta &=& {\Pr}(|\tilde{ P_n} - \langle \hat P_n \rangle|\ge \eta)  \nonumber \\
&=& 2{\Pr}(X \ge \eta\sqrt{M})  \nonumber \\
&=& 2\mathcal{Q}(\eta \sqrt{M}) \le 2e^{-\eta^2M/2}
\end{eqnarray}
where $X = \sqrt{M}(\tilde{ P_n} - \langle \hat P_n \rangle)$ is the standard normal variable. $\mathcal{Q}(x)$ is the Q-function. In the last line we have used the Chernoff bound for Q-function. Inverse the above equation gives: $M \le \frac{2{\rm ln}(10)}{\eta^2}{\rm log}(\frac{2}{\delta})$. If a evolution time $t$ is required for a measurement, then the total evolution time to have ${\Pr}(|\tilde{ P_n} - \langle \hat P_n \rangle |\ge \eta) = \delta$ is $tM = t\frac{2{\rm ln}(10)}{\eta^2}{\rm log}(\frac{2}{\delta}) = \mathcal{O}[t( {\rm log}(\delta^{-1})]$. If $M$ is large, we can assume $\tilde{ P_n} \approx \langle \hat P_n \rangle$, which gives:
\begin{eqnarray}
|\mathcal{Z}(t) - e^{-i\langle \hat P_n \rangle}| &=& |e^{-i\tilde{ P_n}}-e^{-i\langle \hat P_n \rangle}| \nonumber\\
&=& |e^{-i(\tilde{ P_n} - \langle \hat P_n \rangle)}-1| \nonumber \\
&\approx& |-i(\tilde{ P_n}-\langle \hat P_n \rangle)| \nonumber \\
&=&|\tilde{ P_n}-\langle \hat P_n \rangle| \le \eta
\end{eqnarray}
Therefore, generating a signal $\mathcal{Z}(t)$ that satisfies ${\rm Pr}(|\mathcal{Z}(t) - e^{-i\langle \hat P_n \rangle}| \le \eta) = 1-\delta$ would take a total evolution time of $\mathcal{O}[t( {\rm log}(\delta^{-1})]$, which proves the applicability of RFE in the trotter-based scheme. Implementing RFE allows us to obtain ${\Lambda}_{b,l}^n$ at Heisenberg limit.

\section{Solving Hamiltonian coefficients with linear equations}

Now that all $\Lambda_{b, l}^n$ and $\Xi_{b, l}$ are obtained, we wish to find the Hamiltonian coefficients $\lambda_s^n$ and $\xi_s$. As the form of ${\hat E}_b$ is pre-determined, $\gamma_l^s$, which refers to the eigenvalue of $\ket{{\hat E}_{b}}_l$ with respect to ${\hat E}_s$, are also determined. As discussed in the manuscript, $s$ indexes the elements in $S_b$, which goes up to $2^k$ as $|S_b| = 2^k$. Similarly, $l$ indexes the ${\hat E}_b$ eigenstates, which also goes up to $2^k$. For a specific Pauli string ${\hat E}_b$, let $C_l = \Lambda_{b, l}^n$ or $\Xi_{b, l}$, and $c_l = \lambda_s^n$ or $\xi_s$, respectively. Looping over all $s$ and $l$ gives:
\begin{eqnarray}
\begin{bmatrix}
\gamma_{1}^{1} & \gamma_{1}^{2} & ... & \gamma_{1}^{2^k} \\
\gamma_{2}^{1} & \gamma_{2}^{2} & ... & \gamma_{2^k}^{2^k} \\
\vdots & \vdots & \vdots & \vdots\\
\gamma_{2^k}^{1} & \gamma_{2^k}^{2} &  ... & \gamma_{2^k}^{2^k}
\end{bmatrix}
\begin{bmatrix}
c_1 \\
c_2 \\
\vdots \\
c_{2^k}
\end{bmatrix}
=
\begin{bmatrix}
C_1 \\
C_2 \\
\vdots \\
C_{2^k}
\end{bmatrix}.
\end{eqnarray}
We denote the square matrix on the left hand side of the equation to be $\Gamma$, where each row corresponds to an index $s$ and each column corresponds to an index $l$. 

To ensure that the inversion from $C_l$ to $c_l$ does not amplify errors, we here prove that $\Gamma$ is orthogonal up to a factor. Consider the Pauli string $\hat{E}_b = \prod_{i \in \mathrm{supp}({\hat E}_{b})} \mathcal{P}_i^{b}$ with $|{\rm supp}(\hat{E}_b)| = k$, its eigenstate can be written as: 
\begin{eqnarray}
\ket{{\hat E}_{b}}_l = \prod_{i \in \mathrm{supp}({\hat E}_{b})} \ket{\phi}_{i,l}
\end{eqnarray}
where each $\ket{\phi}_{i,l}$ satisfies $\mathcal{P}_i^{b}\ket{\phi}_{i,l} = \eta_{i,l}\ket{\phi}_{i,l}$ with $\eta_{i,l} \in \pm 1$. To formulate the eigenvalues, we introduce a k-bit string $\textbf{l} = (l_1, l_2, l_3 \dots l_k)$, assigning $l_i = 1$ if $\eta_{i,l} = -1$, and $l_i = 0$ when $\eta_{i,l} = 1$. Similarly, for each ${\hat E}_{s}$, which is obtained by swapping arbitrary $\mathcal{P}_i^{b}$ with identity, we introduce another bitstring $\textbf{s} \equiv (s_1, s_2, s_3 \dots s_k)$, where $s_i = 0$ if $\mathcal{P}_i^{b}$ is swapped with identity, and $s_i = 1$ if $\mathcal{P}_i^{b}$ is retained. Then, the elements of $\Gamma$ matrix can be written as: $\gamma_l^s = (-1)^{\textbf{l} \cdot \textbf{s}}$, where $\textbf{l} \cdot \textbf{s} = \sum_j l_js_j$ denotes the bitwise dot product. This structure can be interpreted as follow: at each bit $i$, $\gamma_l^s$ acquires a factor of $-1$ only if $l_i = s_i = 1$, that is, when the local eigenvalue is $\eta_{i,l} = -1$ and $\mathcal{P}_i^{b}$ is not swapped out in $\hat{E}_s$. Recognizing the form of the $k$-bit Hadamard matrix, we observe that $\Gamma$ coincides with the Hadamard matrix up to a normalization factor $\sqrt{2^k}$. Therefore, $\Gamma$ is orthogonal up to this factor, satisfying $\Gamma \Gamma^\top = 2^k \mathcal{I}$. In particular, its condition number satisfies $\kappa(\Gamma) = ||\Gamma||_2 ||\Gamma^{-1}||_2 = 1$, ensuring that the inversion from  $C_l$ to $c_l$ does not amplify errors.

\section{Deviation from effective dynamics}
Due to the finiteness of $R$ in experiment, the actual dynamic deviates from the ideal one due to trotter error and Monte-Carlo error. This deviation is quantitatively described the trace distance between the ideal density matrix and the actual one. $||\Delta \rho||_1$ can be written as:
\begin{eqnarray}
|| \Delta \rho ||_1 &=& || e^{-i\hat{\mathcal H} t }\rho(0)e^{i\hat{\mathcal H} t } - Q_R \rho(0) Q_R^\dagger ||_1 
\end{eqnarray}
where $\rho(0)$ is the density matrix of the system at $t=0$. $Q_R$ represents the unitary sequence inserted during the reshaping process. $\hat{\mathcal H}$ is the effective Hamiltonian. By definition, $e^{-i\hat{\mathcal H}t} = Q_{\infty}$. The form of $Q_R$ varies from learning one term to another. In general, $Q_R$ takes the form of:
\begin{eqnarray}
Q_R = \prod_{i=1}^R [\mathbb{U}^{\rm left}(\boldsymbol \theta_i)]^\dagger e^{-i{\hat H}\tau}\mathbb{U}^{\rm right}(\boldsymbol \theta_i)
\end{eqnarray}
where ${\hat H}$ is the general Hamiltonian in Eq.~\ref{sup_genH}. Note that $\mathbb{U}^{\rm left} = \mathbb{U}^{\rm right}$ is true for all parts of our algorithm, except for the trotter-based scheme in learning the spin-boson coupling coefficients, where ${\mathbb U}^{(3)}$ is only included in $\mathbb{U}^{\rm right}$

When $\mathbb{U}^{\rm left} = \mathbb{U}^{\rm right}$, proof process in~\cite{Haoya2023} can be implemented as long as we can prove $|| {\hat H} \ket{\Psi} ||_2 $ and $|| {\hat H}^2 \ket{\Psi} ||_2 $ are bounded. To prove for $|| {\hat H} \ket{\Psi} ||_2 $, we have: 
\begin{eqnarray}
&&|| {\hat H} \ket{\Psi} ||_2 \nonumber \\
&=& ||\Big[{\hat H}_{\rm spin} + {\hat H}_{\rm boson} + {\hat H}_{\rm int} \Big] \ket{\Psi} ||_2 \nonumber \\
&\le&  \sum_a \Big[\xi_a||{\hat E}_a \ket{\Psi_s} ||_2 +  \sum_n^{N_b}\lambda_a^n|| [{\hat E}_a \ket{\Psi_s} ||_2 \cdot ||({\hat b}_n^\dagger + {\hat b}_n) \ket{\Psi_b}||_2 \Big]+ \sum_n^{N_b}\omega_n || {\hat b}_n^\dagger {\hat b}_n  \ket{\Psi_b} ||_2 \nonumber \\
&\le& C_q^k {\rm max}\{\xi_a\} + C_q^kN_b {\rm max}\{\lambda_a^n\} \sqrt{2|\alpha|^2 + 2{\rm Re}(\alpha^2) + 1} + N_b{\rm max}\{ \omega_n \} |\alpha|^2
\end{eqnarray}
Note that we have approximated $\ket{\Psi_{\rm B}}$ as a coherent state. Since in the above equation, all variables scale with $\mathcal{O}(1)$ ($\alpha$ only scales with time in the trotter-based scheme, which is not considered here), we conclude $|| {\hat H} \ket{\Psi} ||_2$ is bounded. Similarly, for $|| {\hat H}^2 \ket{\Psi} ||_2$, the number of term is squared in comparison to $|| {\hat H} \ket{\Psi} ||_2$. However, the norm if each terms remains bounded. Therefore, $|| {\hat H}^2 \ket{\Psi} ||_2$ remains bounded. Therefore, by applying the same methodology as in~\cite{Haoya2023}, we can establish the upper bound for $||\Delta \rho||_1$, which scales as $\mathcal{O}(t^2/R)$.

\begin{figure}[t]
\centerline{\includegraphics[width=120mm]{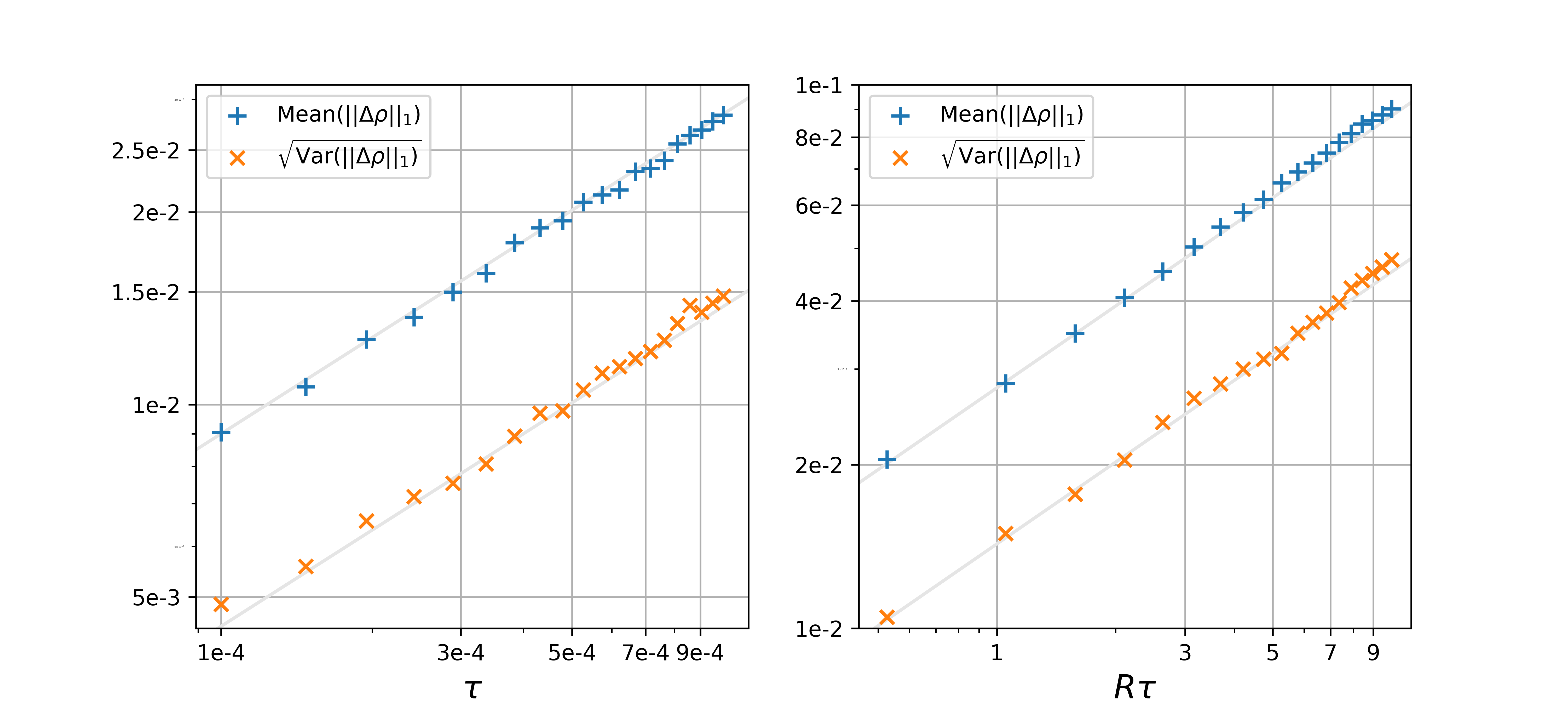}}
\caption{The variance and mean value of trace distance $||\Delta \rho||_1$ are plotted against $\tau$ and $R\tau = t$}
\label{trace_distance}
\end{figure}

However, the above proof fails for the trotter-based scheme in learning spin-boson coupling terms, where $\mathbb{U}^{\rm left} = \mathbb{U}^{\rm right}$. The failure can be attributed to two reasons: 1. The ideal bosonic state is a coherent state with $\alpha \sim \mathcal{O}(t)$. 2. Insering of $\mathbb{U}^{(3)}$ introduces a constant error in the first-order term of $||\Delta \rho||_1$, which fails to prove $||\Delta \rho||_1 \rightarrow 0$ when $R \rightarrow \infty$ using the above technique. Therefore, to quantify the deviation scaling in this case, we numerically plot ${\rm Mean}[|| \Delta \rho ||_1]$ and ${\rm Var}[|| \Delta \rho ||_1]$ with respect to $\tau$ and $t$ in Fig.~\ref{trace_distance}. 1000 sets of $Q_R$ are generated for each subplots to ensure convergence. In the left subplot, $t$ is held constant while $\tau$ is choosen from $10^{-4}$ to  $10^{-3}$. Similarly, in the right subplot, $\tau$ is kept at $10^{-3}$ while $t$ is chosen from $0$ to $10$.  It is observed that ${\rm Mean}[|| \Delta \rho ||_1] \propto \sqrt{{\rm Var}[|| \Delta \rho ||_1]} \propto \sqrt{t}$ at constant $\tau$, and ${\rm Mean}[|| \Delta \rho ||_1] \propto \sqrt{{\rm Var}[|| \Delta \rho ||_1]} \propto \sqrt{\tau}$ at constant $t$. Therefore, the following conclusion can be drawn:
 \begin{eqnarray}
{\rm Mean}[|| \Delta \rho ||_1] \propto \sqrt{{\rm Var}[|| \Delta \rho ||_1]}   \propto \sqrt{t\tau} = t/\sqrt{R}
 \end{eqnarray}

 \section{Derivation for Eq.~\ref{displacement}}
Here, we provide detailed derivation for Eq.~\ref{displacement} in the manuscript. The form of $\hat{\mathcal H}_{\rm SB}^{(2)}$ can be found from Eq.~\ref{Eff_Spin_H1}. If the wavefunction is initialized on $\ket{{\hat E}_b}_l\otimes\ket{\boldsymbol 0}$, the effective time evolution can be written as: 
\begin{eqnarray}
\ket{\Psi_{\rm SB}(t)}_l &=& e^{-i \hat{\mathcal H}_{\rm SB}^{(2)}t} \ket{{\hat E}_b}_l\ket{\boldsymbol 0} \nonumber \\
&=& e^{-i\Xi_{b,l}t}\prod_n e^{-i[\Lambda_{b,l}^n(\hat{b}_n^\dagger + \hat{b}_n) + \omega_n\hat{b}_n^\dagger\hat{b}_n ]t} \ket{{\hat E}_b}_l\ket{\boldsymbol 0}
\end{eqnarray}
where bosonic modes are separated. Omitting the global phase $e^{-i\Xi_{b,l}t}$, we denote the effective bosonic Hamiltonian as:
\begin{eqnarray}
\hat{\mathcal H}_{\rm B}^{\rm eff} = \sum_n [\Lambda_{b,l}^n(\hat{b}_n^\dagger + \hat{b}_n) + \omega_n\hat{b}_n^\dagger\hat{b}_n]
\end{eqnarray}
To obtain the analytical formula in Eq.~\ref{displacement}, we explore the resolution of identity and insert $\prod_n{\hat D}_{n}^\dagger(\Lambda_{b,l}^n/\omega_{n}){\hat D}_{n}(\Lambda_{b,l}^n/\omega_{n})$, which yields:
\begin{eqnarray}\label{displacement_transformation}
\ket{\Psi_{\rm B}(t)}_l &=& \prod_n[{\hat D}_{n}^\dagger(\frac{\Lambda_{b,l}^n}{\omega_{n}}){\hat D}_{n}(\frac{\Lambda_{b,l}^n}{\omega_{n}})]e^{-i \hat{\mathcal H}_{\rm B}^{\rm eff}t} \prod_{n^\prime}{\hat D}_{n}^\dagger(\frac{\Lambda_{b,l}^{n}}{\omega_{n}}){\hat D}_{n}(\frac{\Lambda_{b,l}^{n}}{\omega_{n}}) \ket{\boldsymbol 0} \nonumber \\
& = & \prod_n{\hat D}_{n}^\dagger(\frac{\Lambda_{b,l}^n}{\omega_{n}}) e^{-i \hat{\mathcal H}^{(n)}_{\rm B}t} \ket{\frac{\Lambda_{b,l}^n}{\omega_{n}}}
\end{eqnarray}
where:
\begin{eqnarray}
\hat{\mathcal H}^{(n)}_{\rm B} &=& {\hat D}_{n}(\frac{\Lambda_{b,l}^n}{\omega_{n}}) \Big[ \Lambda_{b,l}^n({\hat b}_{n}^\dagger + {\hat b}_{n})+ \omega_n {\hat b}_n^\dagger {\hat b}_n \Big] {\hat D}^\dagger_{n}(\frac{\Lambda_{b,l}^n}{\omega_{n}}) \nonumber \\
&=&  -\frac{[\Lambda_{b,l}^n]^2}{\omega_{n}} + \omega_n {\hat b}_n^\dagger {\hat b}_n
\end{eqnarray}
which is a free-field Hamiltonian. As the time evolution of coherent state according to a free-field Hamiltonian is known analytically, starting from Eq.~\ref{displacement_transformation} we have:
\begin{eqnarray}
\ket{\Psi_{\rm B}(t)}_l &=& \prod_n{\hat D}_{n}^\dagger(\frac{\Lambda_{b,l}^n}{\omega_{n}}) e^{-i\hat{\mathcal H}^{(n)}_{\rm B}t} \ket{\frac{\Lambda_{b,l}^n}{\omega_{n}}}  \nonumber \\
&=& \prod_n{\hat D}_{n}^\dagger(\frac{\Lambda_{b,l}^n}{\omega_{n}}) \ket{e^{-i\omega_{n}t}(\frac{\Lambda_{b,l}^n}{\omega_{n}})}  \nonumber \\
&=& \prod_n\ket{\frac{\Lambda_{b,l}^n}{\omega_{n}}(e^{-i\omega_{n}t} - 1)} 
\end{eqnarray}
which produces Eq.~\ref{displacement} in the manuscript. 

In the DQS-based scheme, we initialize the state on the entangled squeezed state $\ket{\Psi(0)} = \ket{\Psi_{\rm B}^{\rm ent}} = \prod_n^{N_b}{\rm exp}[\frac{1}{2}(z^\ast {\hat B}_{n}^2 - z {\hat B}_{n}^{\dagger 2})] \ket{\boldsymbol 0}$. Denote $S_{n}^{(W)}(z) = {\rm exp}[\frac{1}{2}(z^\ast {\hat B}_{n}^2 - z {\hat B}_{n}^{\dagger 2})]$, the time evolution of $\ket{\Psi_{\rm B}^{\rm ent}}$ can be written as:
\begin{eqnarray}\label{Supp_Entangled_displacement}
\ket{\Psi_{\rm B}^{\rm ent}(t)} & = & \prod_n{\hat D}_{n}^\dagger(\frac{\Lambda_{b,l}^n}{\omega_{n}}) e^{-it \hat{\mathcal H}^{(n)}_{\rm B}} {\hat D}_{n}(\frac{\Lambda_{b,l}^n}{\omega_{n}}) S_{n}^{(W)}(z)\ket{\boldsymbol 0} \nonumber \\
& = & \prod_n{\hat D}_{n}^\dagger(\frac{\Lambda_{b,l}^n}{\omega_{n}}) \Big[ e^{-it \hat{\mathcal H}^{(n)}_{\rm B}} {\hat D}_{n}(\frac{\Lambda_{b,l}^n}{\omega_{n}}) e^{it \hat{\mathcal H}^{(n)}_{\rm B}}\Big]  \Big[e^{-it \hat{\mathcal H}^{(n)}_{\rm B}} S_{n}^{(W)}(z) e^{it \hat{\mathcal H}^{(n)}_{\rm B}}\Big]  e^{-it \hat{\mathcal H}^{(n)}_{\rm B}} \ket{\boldsymbol 0} \nonumber \\
& = & \prod_n{\hat D}_{n}^\dagger(\frac{\Lambda_{b,l}^n}{\omega_{n}}) {\hat D}_{n}(e^{-i\omega_{n} t}\frac{\Lambda_{b,l}^n}{\omega_{n}}) S_{n}^{(W)}(e^{-2i\omega_{n} t}z) \ket{\boldsymbol 0} \nonumber \\
& = & \prod_n{\hat D}_{n}(\frac{\Lambda_{b,l}^n}{\omega_{n}}(e^{-i\omega_{n} t} - 1)) S_{n}^{(W)}(e^{-2i\omega_{n} t}z) \ket{\boldsymbol 0}
\end{eqnarray}
In the third line, we have used the following properties:
\begin{eqnarray}
&&e^{-it \hat{\mathcal H}^{(n)}_{\rm B}} S_{n}^{(W)}(z) e^{it \hat{\mathcal H}^{(n)}_{\rm B}} \nonumber \\
&=&  e^{-it \hat{\mathcal H}^{(n)}_{\rm B}}{\rm exp}[\frac{1}{2}(z^\ast {\hat B}_{n}^2 - z {\hat B}_{n}^{\dagger 2})]e^{it \hat{\mathcal H}^{(n)}_{\rm B}} \nonumber \\
&=& {\rm exp}[\frac{1}{2}e^{-it \hat{\mathcal H}^{(n)}_{\rm B}}(z^\ast {\hat B}_{n}^2 - z {\hat B}_{n}^{\dagger 2})e^{it \hat{\mathcal H}^{(n)}_{\rm B}}] \nonumber \\
&=& {\rm exp}[\frac{1}{2}(z^\ast e^{2i\omega_n t}{\hat B}_{n}^2 - z e^{-2i\omega_n t}{\hat B}_{n}^{\dagger 2})] \nonumber \\
&=& S_{n}^{(W)}(e^{-2i\omega_n t}z)
\end{eqnarray}
In Eq.~\ref{Supp_Entangled_displacement}, we have proved that $\ket{\Psi_{\rm B}^{\rm ent}(t)}$ takes the form of a displaced squeezed state that is entangled across all $W$ copies of systems. For the $n^{th}$ mode, at $t = \pi/\omega_{n}$, we have :
\begin{eqnarray}
\ket{\Psi_{{\rm B}, n}^{\rm ent}(\frac{\pi}{\omega_{n}})} = {\hat D}_{n}(-\frac{2\Lambda_{b,l}^n}{\omega_{m}}) S_{n}^{(W)}(z) \ket{\boldsymbol 0}
\end{eqnarray}
Therefore, the effect of $e^{-i\hat{\mathcal H}_{\rm SB}^{(2)}\pi/\omega_{m}}$ is equivalent to a field quadrature displacement operator with $\alpha = -\frac{2\Lambda_{b,l}^n}{\omega_{n}}$, which enables the implementation of DQS in~\cite{Quntao2018} to measure $\Lambda_{b,l}^n$ at Heisenberg limit.

\section{Discretization of spectral density function}
The spectral density function used in the numerical example is:
\begin{eqnarray}\label{sup_Jw}
J(\omega) = \sum_n (\Lambda_n)^2 \delta(\omega-\omega_n) = \frac{\eta \omega}{(\omega^2-\Omega^2)^2 + \gamma^2 \omega^2} \nonumber \\
\end{eqnarray}
To obtain $\Lambda_n$ and $\omega_n$ from $J(\omega)$, we first discretize the frequency domain $[0, \omega_{\rm cut}]$ (we use $\omega_{\rm cut} = 4$ in this work) into $N_b$ intervals $[\omega_{n^\prime}, \omega_{n^\prime +1}]$, where $n^\prime = 0, 1, ..., N_b - 1$. Integrating $J(\omega)$ via a coarse-grain method gives:
\begin{eqnarray}
(\Lambda_n)^2 = \int_{\omega_{k^\prime}}^{\omega_{k^\prime}+1} d\omega J(\omega),~~~~~~~\omega_n = \Big[\int_{\omega_{k^\prime}}^{\omega_{k^\prime}+1} d\omega J(\omega)\omega\Big]/(\Lambda_n)^2
\end{eqnarray}
which allows us to find $\Lambda_n$ and $\omega_n$ numerically.

\section{Derivations for models learnable by our Algorithms}
Here, we provide derivations for the form of Hamiltonian in Table~\ref{tab:models}. For Spin-Peierls model, its Hamiltonian is given in \cite{SPmodel}. For the Holstein model and SSH model, their Hamiltonian are given in fermionic form\cite{Holstein_model, SSH_model}:
\begin{eqnarray}\label{sup_Holstein}
{\hat H}_{\rm hst} = \sum_{\langle i,j\rangle}\xi_{i,j} ({\hat c}_{i}^\dagger {\hat c}_{j} + {\rm H. c.}) + \sum_i \lambda_i ({\hat c}_{i}^\dagger {\hat c}_{i})({\hat b}_i^\dagger + {\hat b}_i) + \sum_i \omega_i {\hat b}_i^\dagger {\hat b}_i
\end{eqnarray}
\begin{eqnarray}\label{sup_SSH}
{\hat H}_{\rm SSH} =  \sum_i \xi_i({\hat c}_{i}^\dagger {\hat c}_{i}) + \sum_{\langle i,j\rangle}\lambda_{i,j} ({\hat c}_{i}^\dagger {\hat c}_{j} + {\rm H. c.})({\hat b}_{i,j}^\dagger + {\hat b}_{i,j})+ \sum_{\langle i,j\rangle}\omega_{i,j}  {\hat b}_{i,j}^\dagger {\hat b}_{i,j}
\end{eqnarray}
By performing Jordan-Wigner transformation, the fermionic Hamiltonian can be effectively transformed to spin. Jordan-Wigner transformation, is given by:
\begin{eqnarray}
{\hat c}_{i}^\dagger = \frac{1}{2}(\prod_{j=1}^{i-1} Z_j)(X_i + i Y_i), ~~~~ {\hat c}_{i} = \frac{1}{2}(\prod_{j=1}^{i-1} Z_j)(X_i - i Y_i)
\end{eqnarray}
Applying the above equation on Eq.~\ref{sup_Holstein} and Eq.~\ref{sup_SSH} gives(Hamiltonian parameters up to coefficients):
\begin{eqnarray}
{\hat H}_{\rm hst} = \sum_{\langle i,j \rangle}\xi_{i,j}(X_i X_j + Y_i Y_j) + \sum_i \lambda_i(1-Z_i)({\hat b}_i^\dagger + {\hat b}_i) + \sum_i \omega_i {\hat b}_i^\dagger {\hat b}_i
\end{eqnarray}
\begin{eqnarray}
{\hat H}_{\rm SSH} =  \sum_i \xi_i(1-Z_i) + \sum_{\langle i,j\rangle}\lambda_{i,j} (X_i X_j + Y_i Y_j)({\hat b}_{i,j}^\dagger + {\hat b}_{i,j})+ \sum_{\langle i,j\rangle}\omega_{i,j}  {\hat b}_{i,j}^\dagger {\hat b}_{i,j}
\end{eqnarray}

Besides the above models, our protocol can also be applied for the lattice Schwinger model, whose Hamiltonian is given by \cite{Schwinger_model}:
\begin{eqnarray}\label{sup_schwinger}
{\hat H}_{\rm swg} = \frac{m}{2} \sum_{n \in \mathbb{Z}} (-1)^n Z_n+ \frac{g\sqrt{x}}{8}\sum_{n \in \mathbb{Z}}[\sigma_n^{+} e^{i\theta(n)}\sigma_n^{-} + {\rm H. c.}]+\sum_{n \in \mathbb{Z}}\frac{1}{2g\sqrt{x}}E(n)^2
\end{eqnarray}
where $\sigma_n^{\pm} = X_n \pm iY_n$. To map the field operators to bosonic operators, we consider a cutoff in electric eigenstate $\ell$, such that electric field values are in the interval $[-\ell, \ell]$. When $\ell$ is large, we can approximate the field operators $E(n)$ and $e^{i\theta(n)}$ with: $E(n) \approx  {\hat b}_{n}^\dagger {\hat b}_{n} - \ell$ and $e^{i\theta(n)} \approx  {\hat b}_{n}^\dagger / \sqrt{\ell}$. This gives:
\begin{eqnarray}
{\hat H}_{\rm swg} = \frac{m}{2} \sum_{n \in \mathbb{Z}} (-1)^n Z_n + \frac{g\sqrt{x/\ell}}{8}\sum_{n \in \mathbb{Z}}[\sigma_n^{+}{\hat b}_n^\dagger\sigma_n^{-} + {\rm H. c.} ]+\frac{1}{2g\sqrt{x}}\sum_{n \in \mathbb{Z}}({\hat b}_{n}^\dagger {\hat b}_{n} - \ell)^2 \nonumber \\
\end{eqnarray}
Note that although ${\hat H}_{\rm swg}$ includes nonlinear terms in the bosonic operators, it remains learnable via the trotter-based scheme, whereas DQS-based approaches would fail due to the nonlinearity. To learn Schwinger model with trotter-based scheme, one should use $\mathbb{U}^{(3)} = {\rm exp}[\frac{1}{2g\sqrt{x}}\sum_{n \in \mathbb{Z}}({\hat b}_{n}^\dagger {\hat b}_{n} - \ell)^2]$. After inserting $\mathbb{U}^{(3)}$, one should measure $\langle {\hat X_n} \rangle$ and $\langle {\hat P_n} \rangle$ simultaneously, as for Schwinger model ${\hat H}_{\rm int}$ also carries ${\hat P_n}$.

\section{Pseudocode for Hamiltonian learning of hybrid quantum systems}
To present a more accessible description of the protocol introduced in the manuscript, we restate the algorithm in the format of pseudocode. Five functions are used in the pseudocode below: $\texttt{RUT}$ takes a Hamiltonian and a unitary sequence as inputs and return the reshaped Hamiltonian according to Eq.~\ref{sup_insert}. $\texttt{RPE}$, $\texttt{RFE}$ and $\texttt{DQS}$ takes a Hamiltonian (or a displacement channel), an initial state, and a target accuracy as inputs and return the learnt parameters. Readers are referred to ~\cite{Shelby2015}, ~\cite{Haoya2023}, and ~\cite{Quntao2018} for more information. $\texttt{SolveLE}$ takes a column array and a coefficient array and return the solution to the corresponding linear equation.

\begin{algorithm}
\caption{Learning pure spin and boson coefficients}
\KwIn{Unknown Hamiltonian $\hat H$, $k$, $N_q$, $N_b$, and target accuracy $\epsilon$
}
\KwOut{Estimation ${\tilde \xi}_a$, ${\tilde \omega}_n$}
Eb\_list $\leftarrow$ [All possible $\hat E_b$ with $|{\rm supp}(\hat {\hat E}_b)| = k$]\\
Construct $\mathbb{U}^{(1)}$ \\
$\hat{\mathcal H}^{(1)}$ $\leftarrow$ RUT($\hat H$, $\mathbb{U}^{(1)}$) \\
b\_states $\leftarrow$ $\prod_n D_n(\alpha) \ket{\boldsymbol{0}}$  \\
\For{$n$ \textbf{in} $1\colon N_b$}{
    ${\tilde \omega}_n$ $\leftarrow$ RFE($\hat{\mathcal H}^{(1)}$, b\_states, $\epsilon$)\\
}
\For{${\hat E}_b$ \textbf{in} \rm{Eb\_list}}{
    Construct $\mathbb{U}^{(2)}_b$ \\ Construct $S_{b}$\\
    $\hat{\mathcal H}^{(2)}_{\rm S}$ $\leftarrow$ RUT($\hat{\mathcal H}^{(1)}$, $\mathbb{U}^{(2)}_b$)\\
    Es\_list $\leftarrow$ [All elements in $S_b$] \\
    El\_list  $\leftarrow$ [All possible eigenstates of ${\hat E}_b$] \\
    ElPair\_list  $\leftarrow$ [All possible pairs of $\ket{E_b}_l$] \\
    \For{\rm{$\ket{E_b}_l$} \textbf{in} \rm{El\_list}}{
        \For{${\hat E}_s$ \textbf{in} \rm{Es\_list}}{
            Gamma\_list $\leftarrow$ $\gamma_l^s$ such that ${\hat E}_s\ket{E_{b}}_l = \gamma_l^s \ket{E_{b}}_l$ \\
        }

    }
    \For{\rm{ElPair} \textbf{in} \rm{ElPair\_list}}{
        s\_state $\leftarrow$  construct initial states with Eq.~\ref{sup_RPE_state}\\
        Phase\_diff $\leftarrow$ RPE($\hat{\mathcal H}_{\rm S}^{(2)}$, s\_state, $\epsilon$)\\
        coeff\_list$\leftarrow$ [coefficients of ${\tilde \Xi}_{b,l}$ in Phase\_diff]\\
    }
    ${\tilde \Xi}_{b,l}$ $\leftarrow$ SolveLE(Phase\_list, coeff\_list) \\
    ${\tilde \xi}_a$ $\leftarrow$ SolveLE(All ${\tilde \Xi}_{b,l}$, Gamma\_list)\\

}
\label{algo:pure}
\end{algorithm}

\begin{algorithm}
\caption{trotter-based scheme for spin-boson coupling coefficients}
\KwIn{Unknown Hamiltonian $\hat H$, $k$, $N_q$, $N_b$, ${\tilde \omega}_n$ and target accuracy $\epsilon$}
\KwOut{Estimation ${\tilde \lambda}_a^n$}
Eb\_list $\leftarrow$ [All possible $\hat E_b$ with $|{\rm supp}(\hat {\hat E}_b)| = k$]\\
\For{${\hat E}_b$ \textbf{in} \rm{Eb\_list}}{
    Construct $\mathbb{U}^{(2)}_b$ \\
    Construct $S_{b}$\\
    $\hat{\mathcal H}_{\rm SB}^{(2)}$ $\leftarrow$ RUT($\hat H$, $\mathbb{U}^{(2)}$) \\
    Es\_list $\leftarrow$ [All elements in $S_b$] \\
    El\_list  $\leftarrow$ [All possible eigenstates of ${\hat E}_b$] \\
    \For{\rm{$\ket{E_b}_l$} \textbf{in} \rm{El\_list}}{
        \For{${\hat E}_s$ \textbf{in} \rm{Es\_list}}{
            Gamma\_list $\leftarrow$ $\gamma_l^s$ such that ${\hat E}_s\ket{E_{b}}_l = \gamma_l^s \ket{E_{b}}_l$ \\
        }

    }
    \For{\rm{$\ket{E_b}_l$} \textbf{in} \rm{El\_list}}{
        $\hat{\mathcal H}^{(n)}_{\rm B}$ $\leftarrow$ Insert $\mathbb{U}^{(3)}$ following Eq.~\ref{insert}\\
        b\_states $\leftarrow$ $\ket{\boldsymbol{0}} = \ket{\rm vac}^{\otimes N_b}$ \\
        \For{$n$ \textbf{in} $1\colon N_b$}{
            $\Lambda_{b,l}^n$ $\leftarrow$ RFE($\hat{\mathcal H}^{(n)}_{\rm B}$, b\_states, $\epsilon$)\\
        }
    }
    ${\tilde \lambda}_a^n$ $\leftarrow$ SolveLE($\Lambda_{b,l}^n$, Gamma\_list)\\
}
    \label{algo:trotter}
\end{algorithm}

\begin{algorithm}
\caption{DQS-based scheme for spin-boson coupling coefficients}
\KwIn{Unknown Hamiltonian $\hat H$, $k$, $N_q$, $N_b$, ${\tilde \omega}_n$, $z$, $W$ and target accuracy $\epsilon$}
\KwOut{Estimation ${\tilde \lambda}_a^n$}
Eb\_list $\leftarrow$ [All possible $\hat E_b$ with $|{\rm supp}(\hat {\hat E}_b)| = k$]\\
\For{${\hat E}_b$ \textbf{in} \rm{Eb\_list}}{
    Construct $\mathbb{U}^{(2)}_b$ \\
    Construct $S_{b}$\\
    $\hat{\mathcal H}_{\rm SB}^{(2)}$ $\leftarrow$ RUT($\hat H$, $\mathbb{U}^{(2)}$) \\
    Es\_list $\leftarrow$ [All elements in $S_b$] \\
    El\_list  $\leftarrow$ [All possible eigenstates of ${\hat E}_b$] \\
    \For{\rm{$\ket{E_b}_l$} \textbf{in} \rm{El\_list}}{
        \For{${\hat E}_s$ \textbf{in} \rm{Es\_list}}{
            Gamma\_list $\leftarrow$ $\gamma_l^s$ such that ${\hat E}_s\ket{E_{b}}_l = \gamma_l^s \ket{E_{b}}_l$ \\
        }

    }
    \For{\rm{$\ket{E_b}_l$} \textbf{in} \rm{El\_list}}{
        b\_states $\leftarrow$ construct an entangled state with $z$ and $W$ from Eq.~\ref{multimode_entangled} \\
        \For{$n$ \textbf{in} $1\colon N_b$}{
            $t^\ast$ $\leftarrow$ $\pi/ {\tilde \omega}_n$ \\
            $D$ $\leftarrow$ $e^{-i\hat{\mathcal H}_{\rm SB}^{(2)} t^\ast}$\\
            $\Lambda_{b,l}^n$ $\leftarrow$ DQS($D$, b\_states, $\epsilon$)\\

        }
    }
    ${\tilde \lambda}_a^n$ $\leftarrow$ SolveLE($\Lambda_{b,l}^n$, Gamma\_list)\\
}
    
\label{algo:trotter}
\end{algorithm}

\end{document}